\documentclass[11pt]{article}
\usepackage{jheppub}
\usepackage{amsmath,amssymb,amsthm,mathtools,bm,mathrsfs}
\usepackage{bold-extra}
\usepackage[dvipsnames]{xcolor}
\usepackage{comment}
\usepackage{hyperref}
\usepackage[utf8]{inputenc}
\usepackage[titletoc]{appendix}
\usepackage{tikz}
\usetikzlibrary{shapes,arrows,shadows}
\usepackage[colorinlistoftodos]{todonotes}
\usepackage{xspace}
\usepackage{etoolbox}
\usepackage{listofitems}
\usepackage{tikz-cd}
\usepackage{physics}
\usepackage[font={small,sf}]{caption, subcaption}
\usepackage{booktabs}
\usepackage{adjustbox}
\usepackage{float}

\makeatletter
\def\@fpheader{\relax}
\makeatother

\newcommand\be{\begin{equation}}
\newcommand\ee{\end{equation}}
\newcommand\bea{\begin{eqnarray}}
\newcommand\eea{\end{eqnarray}}
\newcommand\ba{\begin{array}}
\newcommand\ea{\end{array}}

\newcommand\bc{\begin{center}}
\newcommand\ec{\end{center}}

\renewcommand\comment[1]{}
\usepackage{blindtext}
\usepackage{enumitem}

\renewcommand\tilde{\widetilde}

\def\IC{\mathbb{C}}

\definecolor{alertred}{HTML}{BF0603}

\DeclareMathOperator*{\argmin}{argmin}

\newtheorem{thm}{Theorem}

\newcounter{descriptcount}
\newlist{enumdescript}{description}{1}
\setlist[enumdescript,1]{%
  before={\setcounter{descriptcount}{0}%
          \renewcommand*\thedescriptcount{\arabic{descriptcount}}},
        font={\bfseries\stepcounter{descriptcount} \thedescriptcount.~}
}

\newtheoremstyle{indented}{3pt}{3pt}{\addtolength{\leftskip}{2.5em}}{}{\bfseries}{.}{.5em}{}

\theoremstyle{indented}

\newcommand{\mc}[1]{\mathcal{#1}}

 \allowdisplaybreaks
 \numberwithin{equation}{section}

\title{Hermitian Yang--Mills connections on general vector bundles: geometry and physical Yukawa couplings}

\author{Challenger Mishra}
\author{\!\!, Justin Tan}

\affiliation{Department of Computer Science \& Technology, University of Cambridge, Cambridge CB3 0FD, UK}

\emailAdd{cm2099@cam.ac.uk}
\emailAdd{jt796@cam.ac.uk}

\abstract{
We compute solutions to the Hermitian Yang--Mills equations on holomorphic vector bundles $V$ via an alternating optimisation procedure founded on geometric machine learning. The proposed method is fully general with respect to the rank and structure group of $V$, requiring only the ability to enumerate a basis of global sections for a given bundle. This enables us to compute the physically normalised Yukawa couplings in a broad class of heterotic string compactifications. Using this method, we carry out this computation in full for a heterotic compactification incorporating a gauge bundle with non--Abelian structure group.
}

\begin{document}

\maketitle

\section{Background}
Our goal is to solve the Hermitian Yang--Mills (HYM) equations \cite{Siu1987} for a holomorphic vector bundle over a K\"ahler manifold, $V \rightarrow (X, \omega)$. Let $F$ be the curvature of the Chern connection on $V$ endowed with a Hermitian structure $H$. The HYM equations state the contraction of the curvature with the K\"ahler form is proportional to the identity on the bundle fibres;\footnote{Our mathematical conventions are summarised in Appendix \ref{app:conventions}.}
\begin{equation}\label{eq:hym}
 \Lambda F = \omega^{\overline{\nu} \mu} F_{a \phantom{b} \mu \overline{\nu}}^{\phantom{a} b} = \lambda \mathbf{1}_V~.
\end{equation}
Here the Einstein constant $\lambda$ is some real global constant purely depending on the topology of $V \rightarrow X$ via \eqref{eq:slope_conventions}. This is a second order elliptic partial differential equation for the Hermitian bundle metric $H$. To the best of our knowledge, an explicit form for $H$ is not known for any non--trivial $V \rightarrow X$ with non--Abelian structure group, save for $T\mathbb{P}^n$. For stable $\textsf{SU}(n)$ bundles, which are of physical interest in string compactifications, $\det V$ is globally trivialisable by definition; the first Chern class vanishes, $c_1(V)= 0$, and so does the slope, $\lambda = 0$. Our work will be fully general \textit{w.r.t.} the structure group and rank of $V$. Before beginning, we will need to recall the following correspondence between slope--stability and canonical metrics on $V$ \cite{donaldson_hym, uhlenbeck_yau_hym},
\begin{thm}{(Donaldson--Uhlenbeck--Yau)}
    A holomorphic vector bundle $V \rightarrow (X, \omega)$ over a compact K\"ahler manifold admits a unique HYM connection if and only if it is slope poly--stable.
\end{thm}
The Donaldson--Uhlenbeck--Yau theorem guarantees existence and uniqueness of the HYM connection, but is non--constructive, as with the Calabi--Yau theorem. It is therefore natural to turn to numerical approximations of the HYM connection on $V$. 

The physical motivation for studying \eqref{eq:hym} arises from finding a compactification background for the heterotic string which preserves some degree of supersymmetry. Let $F$ denote the gauge field strength of the Yang--Mills fields coupled to supergravity, associated with some connection on a vector bundle over a Calabi--Yau manifold $V \rightarrow X$. The condition for the vanishing of the ten--dimensional gaugino variation under a supersymmetry transformation along the directions tangent to the compact space $X$ results in the system \cite{Green:1987mn}
\begin{equation}
    F^{(0,2)} = F^{(2,0)} = \Lambda F^{(1,1)} = 0~.
\end{equation}
The first two equations imply the connection is compatible with the complex structure, making $V$ a holomorphic vector bundle. The last condition is precisely the HYM equation \eqref{eq:hym} with zero Einstein constant. Obtaining the HYM connection is a prerequisite for computing quantitative details of the four--dimensional physics obtained by dimensional reduction of the heterotic string. Concretely, this requires us to find zero modes of the Dolbeault Laplacian on $(0,1)$ $V$--valued forms;
\begin{equation}\label{eq:zero_modes_dolbeault}
\textsf{ker}\left(\Delta_V = \bar{\partial}_V^{\dagger} \bar{\partial}_V + \bar{\partial}_V \bar{\partial}_V^{\dagger}: \Lambda^{0,1} \otimes V \rightarrow \Lambda^{0,1} \otimes V \right)~.
\end{equation}
The presence of the codifferential in $\Delta_V$ means that finding the zero modes \eqref{eq:zero_modes_dolbeault} requires explicit knowledge of the unique HYM connection $\nabla$ on $V$. The difficulty of finding an exact solution to the Hermitian Yang--Mills system \eqref{eq:hym} and the importance of such solutions to string phenomenology has motivated the numerical study of this system. The earliest attempt at numerically approximating HYM connections \cite{Douglas:2006hz} is based on an iterative method. It may be regarded as a generalisation of Donaldson's algorithm for obtaining canonical metrics on the holomorphic line bundle $\mc{L}^k \rightarrow X$ using sections which generate the embedding into projective space $\iota: X \hookrightarrow \mathbb{P}^n$. The core idea of the generalisation is to find a canonical Hermitian structure on a twist of a stable holomorphic vector bundle $V$ using a collection of global sections defined by the Kodaira embedding of $X$ into a suitable Grassmannian;
\begin{equation}
    \iota: X \hookrightarrow \mathsf{Gr}\left(\rank V; \dim H^0(X;V \otimes \mc{L}^k)\right)~.
\end{equation}
The Hermitian structure on $V$ may then be expanded in a basis of such sections, and converges to the unique solution of \eqref{eq:hym} under an iterative procedure first proposed in \cite{wang2005}. Subsequent work applied the generalised Donaldson algorithm to study the stability properties of a wide class of vector bundles, providing a numerical criterion for determining the existence of supersymmetric vacua in heterotic compactification \cite{Anderson:2010ke, Anderson:2011ed}. The generalised Donaldson algorithm was fruitfully applied to a general class of vector bundles obtained via the \textbf{monad construction} over projective varieties (Section \ref{sec:examples}). These give a large class of non--Abelian bundles over Calabi--Yau $n$--folds realised as projective varieties. 

More recently, machine learning methods have been used to approximate HYM connections, initiated in \cite{Ashmore:2021rlc}, which uses a neural network ansatz to model the desired Hermitian structure. Subsequent work along these lines applied machine learning methods to compute phenomenological information from certain string compactification models \cite{Larfors:2022nep, Constantin:2024yxh}. The main limitation of these methods is that they are restricted to vector bundles constructed as the Whitney sum of holomorphic line bundles over a Calabi--Yau $X$ embedded in a product of projective spaces $\mc{A} = \mathbb{P}^{n_1} \times \cdots \times \mathbb{P}^{n_K}$,
\begin{equation}
    V_L = \bigoplus_{a=1}^n \mc{O}_X(\textbf{k}_a)~.
\end{equation}
Where $\mc{O}_X(\mathbf{k}) \rightarrow X$ denotes a line bundle over the Calabi--Yau embedded in $\mc{A}$. Such bundles have an Abelian structure group $S(U(1)^n)$ and constitute a small fraction of known bundle constructions in string compactifications. In this work, we propose a general machine learning method for approximating HYM connections on holomorphic vector bundles $V$ of arbitrary rank and non--Abelian structure group. The only prerequisite for our approach is that one should be able to explicitly write down a basis of sections of $V$, or of an appropriate twisting $V \otimes \mc{L}^k$ for $k$ sufficiently large and $\mc{L}$ a holomorphic line bundle over $X$.

\textbf{Yukawa couplings} are parameters appearing in quantum field theories which describe how strongly observable particles interact. In some sense string theory acts as a high--dimensional avatar for observable physics in the sense that the geometry and topology of the bundle $V$ and compactification manifold $X$ dictate the field content of the low--energy physical theory and the manner of their interactions. In Section \ref{sec:physical_yukawa} we will go the full mile --- armed with the Ricci--flat metric $g$ on $X$ and the HYM connection $\nabla$ on $V$, we will exhibit a general computation of the physical Yukawa couplings for a non--Abelian bundle $V$.

\section{Approach}
Many interesting problems in differential geometry may be phrased as finding the `optimal' representative in a given cohomology class. These will exist and are unique for the systems we are interested in. We will follow this approach to find the unique HYM connection on a stable bundle $V$. The curvature associated with the Chern connection $F$ is a $\bar{\partial}$--closed $(1,1)$--form taking values in $\text{End}(V)$, the endomorphism bundle of $V$ over $X$. $F$ is a representative of the Dolbeault cohomology class, $[F] \in H^1(X; T^*_X \otimes \text{End}(V))$. This is independent of the choice of Chern connection on $V$, \textit{i.e.} $[F^{\nabla + a}] = [F^{\nabla}] = A(V)$ for $a \in \Lambda^{1,0}(\text{End}(V))$. Here $A(V)$ is the Atiyah class of the bundle \cite{atiyah_class, Anderson:2011ty} --- the unique cohomology class in $H^1(X; T^*_X \otimes \text{End}(V))$ in which the curvature form associated to any Chern connection resides. Starting with some reference Chern connection $\nabla_0$ on $V$ associated to the Hermitian structure $H_0$, we will search for the unique HYM connection on $V$ via a deformation of the Hermitian structure of $V$; $H = h \cdot H_0$, where $h$ is a global section of $\text{End}(V)$. This results in an affine shift in the connection, $\nabla \mapsto \nabla + a$, which in turn induces a $\bar{\partial}$--exact correction to the curvature induced by the reference connection,
\begin{equation}\label{eq:F_ansatz}
    F^{\nabla} = F^{\nabla_0} + \bar{\partial} a~, \quad a \in \Lambda^{1,0}(\text{End}(V))~.
\end{equation}
This manifestly preserves compatibility with the holomorphic structure of $V$. In this sense, the curvature of the HYM connection is the `optimal' representative within the Atiyah class $A(V)$, where we will define our notion of optimality below. With this ansatz for the curvature in hand, our general procedure is outlined below:
\begin{enumerate}
    \item Via some geometric ansatz, reduce the problem to finding a vector--valued global function $\{u : u_i \in C^{\infty}(X)\}$, s.t. the HYM condition \eqref{eq:hym} is locally satisfied.
    \item Develop a variational formulation such that finding a solution to the HYM equations is equivalent to minimisation of an objective functional over some function class $\mathcal{U}$ containing the true solution,
    \begin{equation}\label{eq:var_schematic}
    \Lambda F = \lambda \mathbf{1}_V \iff \min_{u \in \mathcal{U}} \mathscr{L}[u]~.
    \end{equation}
    The functional $\mathscr{L}$ should be manifestly coordinate--invariant, otherwise it would depend on the arbitrary choice of coordinates over $V \rightarrow X$.
    \item Discretise the problem in function space by parameterising $u$ through the approximant $u_{\theta}$. The variational objective is minimised in the parameter space of the restricted function class,
    \[ \theta^*  := \argmin_{\theta \in \Theta} \mathscr{L}(\cdots; \theta)~.\]
\end{enumerate}
Step \textit{(3)} will be achieved using a neural network as an ansatz for the global function $u$, with the appropriate mathematical properties encoded into the structure of the network, as will shortly be discussed. The true solution will not lie in this discretised function space, but `correctness' of the solution should be an `open condition' --- in the sense that hypotheses sharing similar values of the variational objective $\mathcal{L}$ should exhibit similar macroscopic properties. Assuming this is true, hypotheses close to the optimum \eqref{eq:var_schematic} may be used as a substitute for the true solutions in subsequent computations. Previous work concerned with finding optimal Riemannian metrics on $X$ has found numerical evidence that this is indeed the case \cite{Butbaia:2024tje, Berglund:2024uqv}. 

We now discuss how to achieve step \textit{(1)}. To begin, we justify why we work in the space of Hermitian endomorphisms of $V$. Recall that the Chern connection $\nabla$ with respect to a given Hermitian structure is determined uniquely by a Dolbeault operator $\nabla^{0,1} = \bar{\partial}$. Explicitly, one has $\nabla = (\partial + \partial H H^{-1}) + \bar{\partial}$. Such connections are in one--to--one correspondence with Hermitian structures on the underlying vector bundle $V$, and we may think interchangeably about the two spaces. Using this association, it is more convenient to work with the space of Hermitian structures as opposed to the space of connections, which is an affine space over the infinite--dimensional real vector space $\Lambda^1 \otimes \text{End}(V)$. If working directly with the space of connections, one would have to continuously project onto the subset of connections compatible with the holomorphic structure at each step, which amounts to ensuring the integrability condition $(F^{\nabla})^{(2,0)} = (F^{\nabla})^{(0,2)} = 0$ for our hypothesised connection $\nabla$ is satisfied at each step. 

Next we will need to recall the relationship between any pair of Hermitian metrics $(H,H_0)$ on $V$ related by some smooth endomorphism $h: H_0 \mapsto  h \cdot H_0 $, which is Hermitian \textit{w.r.t.} both metrics
\begin{equation}\label{eq:curvature_correction}
    F^{\nabla} = F^{\nabla_0} + \bar{\partial}\left((\partial_{H_0}h) h^{-1}\right)~.
\end{equation}
Here $\partial_H$ denotes the $(1,0)$ part of the covariant derivative with respect to the Chern connection associated to $H$. Motivated by \eqref{eq:curvature_correction}, our ansatz will parameterise an endomorphism $h \in \text{End}(V)$ representing a deformation of some background Hermitian structure $H_0$. This reduces our hypothesis space for the HYM connection from the set of all Hermitian structures on $V$ to the more tractable space of the convex cone of Hermitian endomorphisms of $V$ which are positive--definite \textit{w.r.t.} the background Hermitian structure $H_0$.

We will need to expand $h$ in a basis of sections to ensure our approximation procedure is fully covariant. As mentioned before, for the bundles we are interested in, $H^0(X;V) = 0$, and we will hereafter work on the twisted bundle $V(k) = V \otimes \mc{L}^k$, which will have a basis of sections for $k$ sufficiently large by the Kodaira embedding theorem \eqref{eq:k_embedding}\footnote{We review the necessary background on the Kodaira embedding in Appendix \ref{app:embedding}.}. The background metric $H_0$ may be taken to be the generalised Fubini--Study metric on the twisted bundle $V(k)$ given by taking the Hermitian form \eqref{eq:H_form} to be the identity on $H^0(X; V \otimes \mc{L}^k)$. This is the natural analogue of the familiar Fubini--Study metric on projective space generalised to the Grassmannian setting. This grants us a representative of the Atiyah class $A(V)$, and we now turn to a discussion of step \textit{(2)} to characterise optimality within this class.

\subsection{Objective functional}\label{sec:objective}
To motivate our variational approach, first recall that a closed $(1,1)$ form $\xi$ is harmonic w.r.t. $(X, \omega)$ if and only if the contraction $\Lambda \xi \in C^{\infty}(X)$ is constant. This is easily seen by noting that $d\xi = 0 \implies \partial \xi = \bar{\partial}\xi = 0$, and applying the K\"ahler identity $[\Lambda, \partial] = i\bar{\partial}^{\dagger}$. Now define $\xi := \Tr F \in H^{1,1}(X; \mathbb{C})$ as the trace of the curvature form over the endomorphism indices. This is closed but not necessarily exact for a non--trivial bundle $V$. Then the curvature $F$ satisfies the HYM condition \eqref{eq:hym} if and only if $\Tr F$ is a harmonic $(1,1)$--form and the contraction of the trace--free part of $F$ vanishes;
\begin{equation}
    \Lambda F = \gamma \mathbf{1} \iff \left\{\Tr F \textrm{ harmonic} \, \land \, \Lambda \left(F - \left(\frac{1}{\text{rank}(V)} \Tr F\right) \mathbf{1}_V\right) = 0 \right\}~.
\end{equation}
By the Donaldson--Uhlenbeck--Yau theorem, a connection satisfying the above conditions exists and is unique for a stable bundle $V$. We seek to construct the HYM connection by starting with a background Hermitian structure $H_0$ and deforming $H_0$ such that the Chern connection of the final metric satisfies the HYM equations \eqref{eq:hym}. We phrase this as an alternating optimisation process targeting the two conditions above:
\begin{itemize}
    \item First we constrain the trace of the colour matrix $\Lambda F$ to the constant value determined by the HYM condition \eqref{eq:hym}. This corresponds to prescribing the curvature of the determinant line bundle $\det V$ such that the Chern form $c_1(\det V) = \frac{i}{2\pi}[\Tr F]$ is harmonic. We refer to this as the `Abelian' part as this is a constraint on the $U(1)$ part of the structure group.
    \item Subsequently, we hold $\det h$ fixed and optimise some covariant ansatz for the endomorphism $h$, described in Section \ref{sec:arch}, to eliminate the contraction of the non--Abelian trace--free part of the curvature with the K\"ahler form.
\end{itemize}
We alternate these two stages until convergence of both objectives.

\subsubsection{Abelian stage}
The first stage seeks to find the Hermitian structure on the determinant line bundle $\det V$ such that the first Chern form $c_1(V) \in H^2(X)$ is harmonic w.r.t. the K\"ahler form $\omega$. That is, we seek a harmonic representative of the cohomology $[\Tr F^{\nabla}] \in H^{1,1}(X)$. On a compact K\"ahler manifold, this can always be done through a conformal change to the bundle metric $H$, thanks to the $\partial \bar{\partial}$--lemma. Accordingly, we make the ansatz $\tilde{H} = e^f H$, where $f \in C^{\infty}(X; \mathbb{R})$ is a learnable global function representing the conformal factor. To a holomorphic bundle equipped with a given Hermitian structure $(V,H)$ we may always associate the top exterior power $\det V := \bigwedge^{\rank V}V$ with corresponding Hermitian structure $\det H$. This is commonly referred to as the determinant line bundle $(\det V, \det H)$. The curvature form on $\det V$ is then given by the familiar expression for line bundles, $F_{\det V} = \partial \bar{\partial} \log \det H = \Tr F_V$. Then our ansatz assumes the form
\begin{align}\label{eq:conformal_factor}
    \eta := \Tr F^{\tilde{\nabla}} &= \partial \bar{\partial} \log \det \tilde{H} = \partial \bar{\partial} \log \det (e^f H) \nonumber\\
    &= \partial \bar{\partial} \log \det H + \partial \bar{\partial} \log \det e^f \nonumber\\
    &= \Tr F^{\nabla} + (\rank V) \partial \bar{\partial} f \in H^{1,1}(X)~, \, f \in C^{\infty}(X, \mathbb{R})
\end{align}
 To find the unique harmonic representative in $[\Tr F^{\nabla}]$, note that $\eta$ is $\bar{\partial}$--closed. Hence, from a geometric perspective, the natural objective function is simply the norm of the codifferential of $\eta$, where we regard the conformal factor $f$ as the relevant degree of freedom;
\begin{equation}
    \mathscr{E}[f] := (\bar{\partial}^{\dagger}\eta,\bar{\partial}^{\dagger}\eta)_{L^2} = \int_X\bar{\partial}^{\dagger}\eta \wedge \overline{\star} \, \bar{\partial}^{\dagger}\eta~.
\end{equation}
As the curvature of $\det V$ is also $\partial$--closed, the harmonicity of $\eta$ is equivalent to $\Lambda \eta = \text{constant}$ by the first K\"ahler identity
\begin{equation}\label{eq:codiff_11}
i \overline{\partial}^{\dagger} = [\partial, \Lambda] \implies (\bar{\partial}^{\dagger} \eta)_{\lambda} = \partial_{\lambda} (g^{\overline{\nu} \mu} \eta_{\mu \overline{\nu}})~.
\end{equation}
As $\eta \in H^{1,1}(X)$ harmonic $\iff \Lambda \eta$ constant, one could also choose to minimise an upper bound $L[\eta]$ on the variance of $\Lambda \eta \in C^{\infty}(X; \mathbb{R})$ using the objective proposed in \cite{Ashmore:2021rlc} for line bundles. Both measures of error are strongly correlated; minimising either of $\Vert \bar{\partial}^{\dagger} \eta \Vert$ or $L[\eta] \geq \mathbf{V}[\Lambda \eta]$ leads to a corresponding decrease in the other quantity, as exhibited in Section \ref{sec:examples}.

\subsubsection{Non--Abelian stage}
In the second stage, we approximate the non--Abelian part of the HYM curvature. We will search for a section of the endomorphism bundle $h$ to deform the conformally modified metric from the Abelian stage, $\tilde{H} = e^f \cdot H_0$;
\begin{equation}\label{eq:endo}
    H = h \cdot \tilde{H} = h \cdot e^f \cdot H_0, \quad h \in \Gamma(X; \text{End}(V))~.
\end{equation}
Here $H_0$ is the background Hermitian structure induced from the Fubini--Study metric on the ambient Grassmannian \eqref{eq:k_embedding} and we hold the conformal factor $f$ fixed. We constrain the determinant $\det h \equiv 1$ pointwise on $X$ to preserve the condition that the representative of the Chern class for $\det V$ is harmonic \eqref{eq:conformal_factor}. We recall the relationship between any pair of Hermitian metrics $(H,H_0)$ on $V$ related by some smooth endomorphism $h: H_0 \mapsto  H_0 \cdot h$, which is Hermitian \textit{w.r.t.} both metrics
\begin{equation}\label{eq:curvature_correction2}
    F^{\nabla} = F^{\nabla_0} + \bar{\partial}\left((\partial_{H_0}h) h^{-1}\right)~.
\end{equation}
The task is to find the representative in $[F^{\nabla_0}]$ with vanishing contraction of the trace--free part with the K\"ahler form via optimisation of some ansatz for $h$. Denote this trace--free part by 
\[ F_0 := F - \left(\frac{1}{\text{rank}(V)} \Tr F\right) \mathbf{1}_V~.\]
Once more, there is a natural geometric objective function for this stage; the $L_2$ norm of $\Lambda F_0$;
\begin{align}\label{eq:ym_energy}
    \mathscr{E}[\nabla] &:= \left(\Lambda F_0^{\nabla}, \Lambda F_0^{\nabla}\right)_{L^2} \\
    &= \int_X \text{dVol}_X \, \text{Tr}\left( \Lambda F_0 \cdot (\Lambda F_0)^{\dagger_H} \right)~, 
\end{align}
where $\dagger_H$ denotes the Hermitian adjoint with respect to the new metric $H$. Because the trace--free part of the curvature remains invariant under twisting, that is, the trace--free parts of $V$ and $V \otimes \mc{L}^k$ coincide, we can carry out this program on any twist of $V$ and translate this back down to $V$, as discussed in Appendix \ref{app:untwisting}. We discuss how to effectively parameterise the endomorphism field \eqref{eq:endo} in the next section.

\subsection{Ansatz structure}\label{sec:arch}
It is crucial that our hypotheses for the curvature of $\det V$ \eqref{eq:conformal_factor} and the curvature of $V$ \eqref{eq:curvature_correction} are globally defined objects. To explain how we can achieve this, we first make some general comments about modelling tensor fields on manifolds. The tensor fields $\mathscr{T}$ that we are modelling numerically have a geometric existence that is independent of their representation in local coordinates. Denote by $\mathscr{T}^{(i)}$ the coordinate representation in patch $U_i \subset X$. Numerical models must unavoidably describe $\mathscr{T}$ via a collection of local fields $\{\mathscr{T}^{(i)}\}_i$. The necessary and sufficient condition for this collection to define a single, globally defined tensor field is that they transform coherently between patches according to the corresponding transition functions $T_{ij}$, which amount to satisfying the relevant cocycle conditions on $U_i \cap U_j$. We will refer to this property as \textbf{global well--definedness}. 
\subsubsection{Equivariance}
In the machine learning literature, this property may be framed in terms of \textbf{equivariance}. Consider a symmetry group $G$ acting on the inputs $p$ to some function $f$ we are parameterising, with a linear action $p \mapsto \rho_g(p)$. The ansatz is said to be equivariant to $G$ if the output transforms in the appropriate representation of the output space,
\begin{equation}\label{eq:equivariant_nn}
    f(\rho_g(p)) = \tilde{\rho}_g\circ f(p)~, \quad g \in G~.
\end{equation}
On a manifold this is not quite the correct picture as the actual group acting on $p \in X$ is the infinite--dimensional diffeomorphism group $\text{Diff}(X)$, but if we restrict ourselves to modelling sections of the tensor bundle via $f$, we will be concerned with the linearisation of the action of an element $T \in \text{Diff}(X)$. For vector fields the relevant group $G$ is the frame bundle of $X$, which is just $\textsf{GL}(n;\mathbb{C})$ on an arbitrary complex manifold. The representation $\tilde{\rho}$ will just be taken to be the fundamental representation, and this amounts to the standard action of the Jacobian on a local presentation of the sections of $TX$. There is a straightforward generalisation to sections of the tensor bundle $\bigotimes_r TX \,\otimes \,\bigotimes_s T^*X$. Constructing an equivariant model to approximate the tensor field $\mathscr{T}$ then amounts to finding a sequence of differentiable operations to model some global function $f \in C^{\infty}(X)$ s.t. \eqref{eq:equivariant_nn} is satisfied. When introducing the bundle $V \rightarrow X$, we must ensure our ans\"atze are equivariant \textit{w.r.t.} transformations between local trivialisations of both the base $X$ and the bundle $V$.

Returning to the problem at hand, provided the ansatz $h \in \Gamma(X; \textsf{End}(V))$ transforms correctly as a section of the endomorphism bundle, the correction term in \eqref{eq:curvature_correction} will be an endomorphism--valued one--form, $(\partial_{H_0}h)h^{-1} \in \Gamma\left(X; \Lambda^{0,1} \otimes \text{End}(V)\right)$, and the hypothesis $F^{\nabla}$ remains an honest curvature form, implying that $[F^{\nabla}]$ is preserved at any point during optimisation. Expanding $h$ in a basis of sections for $V$ and $V^{\vee}$ renders the ansatz manifestly equivariant w.r.t. the $\textsf{GL}(r)$ choice of frame in \eqref{eq:keller_diagram}, recalling the caveat that we are working on the twisted bundle $V \otimes \mc{L}^k$ instead of the original bundle $V$. Given a basis of global sections $\{S^a_m\}$ of $V(k)$, and a choice of frame $\{e_a\}$ for $V$, one may expand our ansatz as
\begin{align}\label{eq:h_ansatz}
    h &= h^a_{\,b} e_a \otimes e^b = h^{mn} (S^a_m e_a) \otimes ((S^{\flat})_{nb}e^b) \\
    &= h^{mn}S^a_m \, (H_0)_{b \overline{c}} \, \bar{S}^{\overline{c}}_n e_a \otimes e^b~. \nonumber
\end{align}
Here $a,b$ denote bundle indices and $m,n$ enumerate the elements of the basis.

\subsubsection{Model architecture}
We require the coefficients $h^{mn}$ to be a Hermitian matrix in order for the endomorphism to be Hermitian with respect to $H_0$. We parameterise $h^{mn}$ using a neural network which outputs a Hermitian matrix of complex--valued global functions on $X$, using the spectral network construction \cite{Berglund:2022gvm} (see a discussion of this point in Appendix \ref{app:global_fn}). The network accepts as input the coordinates $p \in X$, projects the input to a $\mathbb{C}^*$--invariant representation, and constructs a Hermitian matrix via the square--root free Cholesky decomposition, $h = L D L^{\dagger}$, with the output of the spectral network being the lower triangular matrix and diagonal $(L,D)$. The determinant of $h$ is enforced to be one by subtracting the trace from \eqref{eq:h_ansatz} and taking the matrix exponential, noting that $\det \exp(h) = \exp(\tr h)$. The resulting ansatz remains a well--defined section as a section of the endomorphism bundle transforms by conjugation between local trivialisations. Note that the dimension of the coefficient matrix $h^{mn}$ in \eqref{eq:h_ansatz} grows as $\mc{O}(k^{2n})$ for $k$ the twisting degree and $n= \dim_{\mathbb{C}} X$. Computing the curvature form \eqref{eq:curvature_correction2} is morally the same as computing the Hessian of the endomorphism $h$, and becomes prohibitively expensive for a large twisting degree. For monad bundles where a large twisting degree is necessary to generate a global basis of sections, we use a `tensor' product decomposition for the Hermitian coefficient matrix $h^{mn}$, where `tensor' here in a non--mathematical context refers to a multidimensional array over $\mathbb{R}^{m_1} \times \cdots \times \mathbb{R}^{m_S}$,
\begin{equation}\label{eq:tensor_decomposition}
    [h^{mn}] = \sum_{r=1}^R \bigotimes_{s=1}^S F_s^{(r)}~.
\end{equation}
Here the number of product factors $S$ will depend on the structure of the bundle. For example, for bundles defined via the monad construction \eqref{eq:monad_bundle}, $S$ will depend on the multidegree $\mathbf{e}_i$ and the rank of the ambient bundle $E$. As a simple example, an ambient bundle with a single projective space factor leads to the decomposition
\begin{equation}\label{eq:example_tensor_decomp}
    E = \bigoplus_{i=1}^L \mc{O}_X(\ell) \leadsto [h^{mn}] = \sum_r F^{(r)} \otimes \tilde{F}^{(r)}~.
\end{equation}
Here the first factor $F \in \mathbb{C}^{\ell \times \ell}$ captures the interaction between different summands in $E$ and the second factor $\tilde{F}$ is a $\dim H^0(X;V(k)) \times \dim H^0(X;V(k))$ matrix describing interactions between the holomorphic sections of the twisted bundle. The `rank' $R$ of the decomposition \eqref{eq:tensor_decomposition} is a hyperparameter that must be set empirically. In order for $[h^{mn}]$ to be positive--definite, each factor $F_s^{(r)} = L_s^{(r)}D_s^{(r)}(L_s^{(r)})^{\dagger}$ is a Hermitian matrix constructed using the square--root free Cholesky decomposition, with the respective Cholesky pair $(L_s^{(r)},D_s^{(r)})$ output by respective heads of the modified network with a common spectral network backbone. The exact decrease in space complexity is dependent on the size of each factor $F$, but we empirically observed a superquadratic improvement in terms of memory requirements at no observable change to performance for the examples we consider in Section \ref{sec:examples} using the decomposition \eqref{eq:example_tensor_decomp}.

Similarly, the $\partial \bar{\partial}$--potential for the curvature of the determinant bundle \eqref{eq:conformal_factor} is represented by a real--valued spectral network with an independent set of parameters. Taken together, these choices ensure that our approximation procedure is manifestly equivariant with respect to changes in the choice of local trivialisation for both $X$ and $V$\footnote{See Section \ref{app:equivariance} for how one may explicitly numerically verify this.}. The parameters for $f$ and $h$ are optimised in an alternating loop, holding the parameters of $h$ (respectively $f$) fixed. Complete details of the experimental setup may be found in Appendix \ref{app:experiments}.

Empirically, for computationally realistic values of the twisting degree $k$, we are able to attain similar performance working with the class $\mc{H}_{\infty}$ of all smooth Hermitian structures on $V$ as using a hypothesis in $\mc{H}_k$ with significantly higher twisting degree. Recall that the latter class of endomorphisms $\mc{H}_k$ is associated with the generalised Fubini--Study metrics on $V(k)$, in which case $h^{mn}$ would be a constant Hermitian matrix defined by some Hermitian form on $H^0(V(k))$ \eqref{eq:H_form}. While the Tian--Yau--Zeldritch theorem guarantees that constant--coefficient Fubini--Study metrics approximate the HYM connection arbitrarily well \eqref{eq:tyz}, the dimension of the basis $H^0(X;V(k))$ grows as $\mathcal{O}(k^n), \, n = \dim_{\mathbb{C}} X$, making computations at high--$k$ computationally expensive. Working directly with the space of smooth Hermitian endomorphisms allows us to capture the local curvature information on $V$ required to solve the HYM equation \eqref{eq:hym} without incurring the cost of a high--degree embedding. Appendix \ref{eq:k_embedding} contains a more complete discussion of this point.

\subsection{Measures of error}
To evaluate how closely the HYM equations \eqref{eq:hym} are satisfied for a given connection $\nabla$, it is natural to examine the eigenvalues of the contraction 
\begin{equation*}
    \Lambda F \sim \begin{pmatrix}
\lambda_1 & & & \\
& \lambda_2 & & \\
& & \ddots & \\
& & & \lambda_{\text{rk}V}
\end{pmatrix}~.
\end{equation*}
The natural frame--independent error measure \cite{Anderson:2010ke} is simply, using $\Lambda F = \left(F \wedge \omega^{n-1}\right)/\omega^n$,
\begin{equation}\label{eq:err_weak}
    \tau(\nabla) = \frac{1}{2\pi}\frac{1}{\text{Vol}(X) \cdot \rank V} \int_X d\mu_{\Omega} \Tr(\Lambda F^{\nabla}) = \frac{1}{\rank V}\int_X c_1(V) \wedge \omega^{n-1}~.
\end{equation}
On an $\textsf{SU}(n)$ bundle, $\tau(\nabla)$ should vanish if the HYM condition is satisfied by $\nabla$. For a stable bundle, \textit{e.g.} a twisting of an $\textsf{SU}(n)$ bundle by $\mc{L}^k$, $\int_X c_1(V \otimes \mathcal{L}^k) \wedge \omega^{n-1} \in \mathbb{Z}$. However, note $\tau(\nabla)$ is a topological quantity and cannot distinguish between curvature forms in the same cohomology --- the measure \eqref{eq:err_weak} merely diagnoses if the ansatz for the curvature form \eqref{eq:F_ansatz} yields the correct cohomology class for $\Tr F$. Indeed, we find $\tau(\nabla) \approx \tau(\nabla_0)$ modulo integration error for all the examples we consider here. As $\Lambda F^{\nabla}$ should be proportional to the identity matrix on $V$ times a global constant on a stable bundle, a more appropriate error measure for our approach is to consider the integrated variance of the trace $\Tr(\Lambda F)$,
\begin{equation}\label{eq:integrated_variance}
    \mathbf{V}\left[\Tr\Lambda F\right] = \frac{1}{\text{Vol}_X} \int_X d\mu_{\Omega}\, \left(\Tr \Lambda F - \left\langle \Tr \Lambda F\right\rangle \right)^2~.
\end{equation}
    
    
    

\section{Examples: monad constructions}\label{sec:examples}
The examples we will study are stable holomorphic vector bundles with structure group $\textsf{SU}(n)$ over a Calabi--Yau $n$--fold $(X, \omega, \Omega)$, obtained from the monad construction \cite{Anderson:2008uw, Anderson:2010ke}. Monad bundles are known to generate all vector bundles over projective spaces, and give a rich class of bundles over Calabi--Yau $n$--folds realised as projective varieties. These are defined via a short exact sequence of the form
\begin{equation}\label{eq:monad_bundle}
    0 \longrightarrow V \overset{\iota}{\longrightarrow} E \overset{f}{\longrightarrow} Q \longrightarrow 0~.
\end{equation}
Here the bundle $E$ and the quotient $Q$ are given by sums of positive line bundles, 
\begin{equation*}
    E = \bigoplus_i \mc{O}_X(\mathbf{e}_i)~, \quad Q = \bigoplus_j \mc{O}_X(\mathbf{q}_j)~, \quad e_i, q_j \geq 0~.
\end{equation*}
The map $f \in \textsf{Hom}(E,Q)$ may be locally expressed as a matrix of sections $f_{ij}$ with multidegree $\mathbf{q}_j - \mathbf{e}_i$. As $V \cong \ker f$ by exactness, the monad map $f$ encodes the bundle moduli of the compactification. By dualising \eqref{eq:monad_bundle}, we see that the dual bundle $V^{\vee} \cong \text{coker}(f^{\vee})$. This perspective shall be useful later, and we can always interchange between the two using some generalised Fubini--Study metric on $V$ \eqref{eq:fs_ref}. The positivity condition is not strictly necessary and makes no technical difference, calculation--wise, but it is easier to find phenomenologically acceptable bundles by imposing this constraint. For example, positive monads do not give arise to antifamilies by the Kodaira vanishing theorem \cite{Anderson:2010vdj}, \textit{i.e.} $H^1(X;V^{\vee}) = 0$. Monad bundles are especially amenable to computational study as one may readily algorithmically obtain a basis of sections for $V(k)$ using the Kodaira embedding theorem \cite{kodaira_2005} after twisting the bundle by some sufficiently high power of an ample line bundle $\mathcal{L}$. To see this, dualise \eqref{eq:monad_bundle}, twist by $\mc{L}^k$, and pass to the long exact sequence in cohomology;
\begin{align}\label{eq:LES_cohomology}
    0 &\longrightarrow H^0(X; Q^{\vee} \otimes \mc{L}^k) \longrightarrow H^0(X;E^{\vee} \otimes \mc{L}^k) \longrightarrow H^0(X; V^{\vee} \otimes \mc{L}^k) \\
    &\longrightarrow H^1(X;Q^{\vee} \otimes \mc{L}^k) \longrightarrow \cdots \nonumber
\end{align}
$Q^{\vee} \otimes \mc{L}^k = \bigoplus_j \mathcal{O}_X(k - q_j)$ is a positive line bundle for $k$ sufficiently large, hence the curvature form is positive definite in local coordinates, and the Bochner--Kodaira--Nakano identity on a manifold of vanishing Ricci curvature implies that $H^{(0,q)}(X;L) \simeq H^q(X; L)$ vanishes, where $L$ is some positive line bundle. So \eqref{eq:LES_cohomology} yields that the holomorphic sections of $V^{\vee} \otimes \mc{L}^k$ are given by the cokernel
\begin{equation}
    H^0(X; V^{\vee} \otimes \mc{L}^k)\simeq \frac{H^0(X; E^{\vee} \otimes \mc{L}^k)}{f^{\vee}_*H^0(X; Q^{\vee} \otimes \mc{L}^k)}~.
\end{equation}
The requisite computation may be done on $V \otimes \mathcal{L}^k$, then subsequently untwisted to obtain the Hermitian structure on $V$ itself (see discussion in Appendix \ref{app:untwisting}). The monad construction is one of two general bundle construction methods present in the string theory literature, the other being the spectral cover construction for elliptically fibred Calabi--Yau manifolds, which is thought to describe all bundles over this special class.

\subsection{Stable bundle over Fermat quintic}
The first example we shall consider is a rank three stable vector bundle over the Fermat quintic $X \subset \mathbb{P}^4$ (\textit{cf.} \cite{Douglas:2006jp}, \cite{Douglas:2006hz}), defined by the short exact sequence
\begin{equation}\label{eq:stable_bundle}
    0 \longrightarrow \mathcal{O}_X(-1) \overset{f}{\longrightarrow} \bigoplus_{i=1}^4 \mathcal{O}_X \longrightarrow V \longrightarrow 0~.
\end{equation}
The homomorphism $f$ is given by four generic non--intersecting sections of $\mathcal{O}_X(1)$, and we choose $f = (Z_0, \ldots, Z_3)$. For any short exact sequence $0 \rightarrow S \rightarrow E \rightarrow Q \rightarrow 0$, $c_k(E) = \sum_{i+j=k}c_i(S)c_j(Q)$, and the first Chern class $c_1(V) = H$, meaning this bundle has nonzero slope. As $V \simeq \text{coker} f$, the basis elements of the chosen frame $\{e_i\}_{i=1}^4$ for the ambient space $\mathcal{O}_X(k)^{\oplus 4}$ are subject to certain relations in the quotient which define the local frame for $V$. For example, in patch $U_0 = \{Z \in \mathbb{P}^4 : Z_0 \neq 0\}$, one has $e_0 = - \sum_{i=1}^4 (Z_i/Z_0) e_i$. Twist \eqref{eq:stable_bundle} by $\mathcal{O}_X(k)$ to obtain the short exact sequence
\begin{equation}
    0 \longrightarrow H^0(X;\mathcal{O}_X(k-1)) \overset{f}{\longrightarrow}  H^0(X; \mathcal{O}_X(k)^{\oplus 4}) \longrightarrow H^0(X;V(k)) \longrightarrow 0~.
\end{equation}
From this, one obtains an explicit parameterisation for the sections via the quotient for $k \geq 1$;
\begin{equation*}
    H^0(X;V(k)) \simeq \frac{H^0(X; \mathcal{O}_X(k)^{\oplus 4})}{f_* H^0(X;\mathcal{O}_X(k-1))}~.
\end{equation*}
Firstly, we approximate the Ricci--flat Calabi--Yau metric in the lone K\"ahler class $[\omega_0]$ on $X$ defined by restriction of the ambient Fubini--Study form on $\mathbb{P}^4$. This is done at the point $\psi=0$ in moduli space using the ansatz \cite{Larfors:2022nep}
\begin{equation}
    \omega = \omega_0 + i \partial \bar{\partial} \varphi_{\textsf{NN}}~.
\end{equation}
Here $\varphi_{\textsf{NN}}$ is a globally defined function on $X$, parameterised via a `spectral' neural network invariant w.r.t. projective transformations \cite{Berglund:2022gvm}. By the $\partial\bar{\partial}$--lemma, any $\omega \in [\omega_0]$ may be obtained via such an $\partial\bar{\partial}$--exact correction, and we use this to approximate the K\"ahler form of vanishing Ricci curvature. This is achieved by minimisation of the following objective, which vanishes if and only if the Ricci curvature of $(X,\omega)$ does \cite{yau77},
\begin{equation*}
    \ell_{\omega} = \int_X d\mu_{\Omega} \left\vert 1 - \frac{\omega^n}{\Omega \wedge \bar{\Omega}} 
    \right\vert~.
\end{equation*}
Next, working at $k=2$, we parameterise the conformal factor $f$ \eqref{eq:conformal_factor} using the spectral network construction \cite{Berglund:2022gvm} and fix the trace of the curvature form $F$ to the appropriate constant value by minimising the codifferential of $\Tr F$ \eqref{eq:codiff_11} with respect to the parameters of $f$. We alternate this procedure with  minimisation of the non--Abelian part of the curvature \eqref{eq:ym_energy} with respect to the parameters of the endomorphism ansatz \eqref{eq:h_ansatz}. The trajectories of various quantities of interest for this stage are depicted in Figure \ref{fig:quintic_DKLR_opt}. Complete details of the architecture and optimisation procedure are described in Appendix \ref{app:experiments}.

We study how closely our approximation satisfies the HYM condition \eqref{eq:hym}, by evaluating the average value of the function $\Lambda F$ and the associated Monte Carlo error over an independent set of 500,000 points sampled from $X$, and find
\begin{equation}
\left\langle \Lambda F \right\rangle = \frac{1}{\text{Vol}_X}\int d\mu_{\Omega} \, \Lambda F = (3.500 \pm 0.005) \cdot \mathbf{I}_3 \pm \mathcal{O}(10^{-3})~.
\end{equation}
where we compute the variance elementwise as
\[ \mathbf{V}\left[\Lambda F\right] = \frac{1}{\text{Vol}_X} \int_X d\mu_{\Omega} \,\left(\Lambda F - \left\langle \Lambda F \right\rangle \right)^2~, \]
and find $\max \sigma\left(\Lambda F^{\nabla}\right) = 5.48 \times 10^{-3}$ along the diagonal, with the  error for off-diagonal entries being also $\mc{O}(10^{-3})$. We estimate the HYM condition is satisfied post--optimisation within an error of $\left(\max \sigma_{\Lambda F^{\nabla}}\right)/\mu_{\Lambda F^{\nabla}} \approx 0.18\%$, where $\mu_{\Lambda F^{\nabla}} := (\text{Vol}_X)^{-1} \int_X d\mu_{\Omega} \, \text{diag}(\Lambda F)$, the average of the diagonal elements over $X$. This is to be compared with the figure of merit for the background connection, at  $\left(\max \sigma_{\Lambda F^{\nabla_0}}\right)/\mu_{\Lambda F^{\nabla_0}} \approx 11\%$. A compilation of these results may be found in Table \ref{tab:hym_results_DKLR}. Recall the trace $\Lambda F$ is a topological quantity; it is not surprising that this coincides for the background and post--optimisation connections --- what should be noted is that our optimised HYM connection $\nabla$ reduces the variance of the trace by four orders of magnitude relative to the background connection $\nabla_0$. We include plots of quantities of interest over the training procedure in Figure \ref{fig:quintic_DKLR_opt}.

\begin{table}[h]
\centering
\caption{HYM approximation measures over Fermat quintic bundle. Data obtained using commit \href{https://github.com/Justin-Tan/cymyc/commit/63bcd80417c69d224408666f051e330c08c063bb}{63bcd80}.}
\begin{adjustbox}{max width=\textwidth}
\begin{tabular}{lcc}
\toprule
Quantity & Background $\nabla_0$ & Post--optimisation $\nabla$ \\
\midrule
$\langle \Lambda F \rangle$ & 
$\begin{pmatrix}
3.51 & 2.2 \times 10^{-4} & 6.3 \times 10^{-5} \\
5.1 \times 10^{-4} & 3.49 & 5.8 \times 10^{-5} \\
4.8 \times 10^{-5} & 9.1 \times 10^{-5} & 3.50
\end{pmatrix}$ & 
$\begin{pmatrix}
3.50 & -4.6 \times 10^{-6} & -6.3 \times 10^{-5} \\
-1.9 \times 10^{-4} & 3.50 & 5.5 \times 10^{-5} \\
-5.8 \times 10^{-5} & 5.2 \times 10^{-5} & 3.50
\end{pmatrix}$ \\[3ex]
\midrule
Max diagonal $\sigma$ & $0.38$ & $5.5 \times 10^{-3}$ \\
Max off--diagonal $\sigma$ & $0.14$ & $4.7 \times 10^{-3}$ \\
$\langle \Lambda \operatorname{Tr} F \rangle$ & $10.500 \pm 0.93$ & $10.500 \pm 0.006$\\
\bottomrule
\end{tabular}
\end{adjustbox}
\label{tab:hym_results_DKLR}
\end{table}

\begin{figure}
    \centering
    \includegraphics[width=1.0\linewidth]{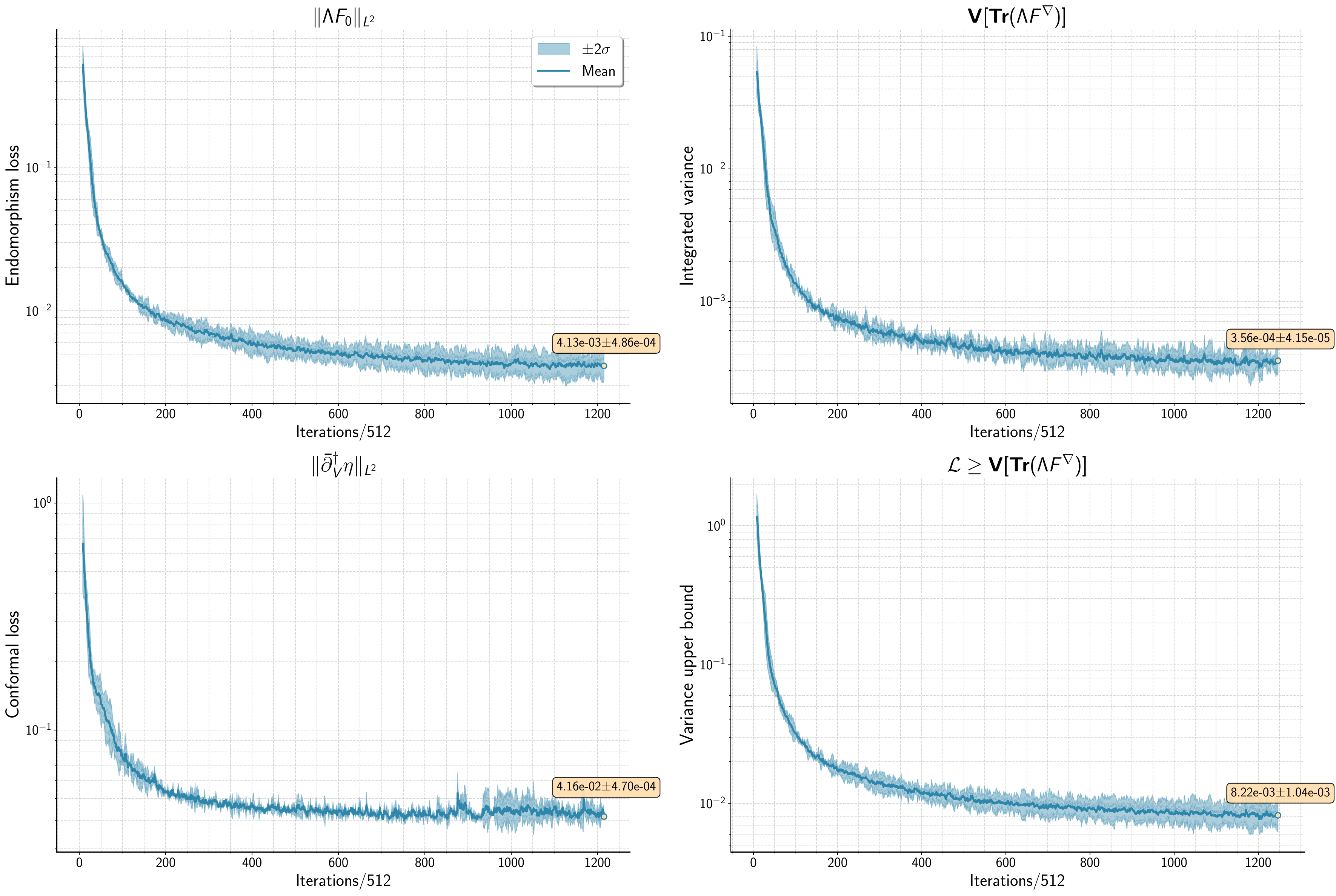}
    \caption{Evolution of various quantities during optimisation for the $c_1=H$ Fermat quintic bundle, evaluated on a separate validation set. Clockwise from top left: \textit{(a)}: Non--abelian objective function; norm of $\Lambda F_0$ \eqref{eq:ym_energy} \textit{(b)}: Integrated variance of $\Tr \Lambda F_0$ \eqref{eq:integrated_variance}, \textit{(c)}: Conformal objective function; codifferential norm \eqref{eq:codiff_11}, \textit{(d)}: Upper bound on the variance of $\Tr \Lambda F_0$ \eqref{eq:var_objective}. Results are reported on three independent runs over the same dataset.}
    \label{fig:quintic_DKLR_opt}
\end{figure}

\subsection{$\textsf{SU}(2)$ bundle over Fermat quartic}
Our second example is an rank two bundle over the Fermat quartic $X \subset \mathbb{P}^3$ defined by the following exact sequence \cite{Anderson:2010ke}
\begin{equation}\label{eq:su2_bundle}
    0 \longrightarrow \mathcal{O}_X(-3) \overset{f}{\longrightarrow} \bigoplus_{i=1}^3 \mc{O}_X(-1) \longrightarrow V \longrightarrow 0~.
\end{equation}
It is apparent from \eqref{eq:su2_bundle} that $c_1(V) = 0$ and hence the bundle has vanishing slope. Here $V \simeq \text{coker} f$, where we fix the bundle moduli to the degree--2 map $f=(Z_0^2, Z_1^2, Z_2^2)$. Twisting \eqref{eq:su2_bundle} appropriately, we obtain a basis of global sections of $V \otimes \mathcal{O}_X(k)$ through the quotient, for $k \geq 3$;
\begin{equation}
    H^0(X; V \otimes \mathcal{O}_X(k)) \simeq \frac{H^0(X; \mathcal{O}_X(k-1)^{\oplus 3})}{f_* H^0(X; \mathcal{O}_X(k-3))}~.
\end{equation}
For a bundle realised as the quotient of some ambient space $\mathcal{O}_X(\mathbf{a})^{\oplus K}$ via the sequence
\begin{equation*}
    0 \longrightarrow S \longrightarrow \mathcal{O}_X(\mathbf{a})^{\oplus K} \longrightarrow V \longrightarrow 0~,
\end{equation*}
for $X \hookrightarrow \mathbb{P}^n$, the number of global sections of the twist $V(k)$ scales as $\dim H^0(X;V(k)) \sim K \cdot {n+k-1 \choose k} - \dim H^0(X;S \otimes \mc{L}^k)$. For this example, $\dim H^0(X;V(k=4)) = 139$. We reuse the Ricci--flat metric computed for the Fermat quintic previously, and compute the HYM connection $\nabla$ for the twisted bundle $V \otimes \mc{L}^4$. Again computing $\langle \Lambda F \rangle$ and $\mathbf{V}[\Lambda F]$ on an independent test set of $500,000$ points on $X$, we find the HYM condition to be satisfied post--optimisation within an error of $(\max \sigma_{\Lambda F^{\nabla}}) / \mu_{\Lambda F^{\nabla}} \approx 0.30\%$. The corresponding figure of merit for the background connection is $(\max \sigma_{\Lambda F^{\nabla}}) / \mu_{\Lambda F^{\nabla}} \approx 18.3\%$, and we summarise the results in Table \ref{tab:hym_results_AG}, with plots of quantities of interest depicted in Figure \ref{fig:quintic_AG_opt}.

\begin{table}[h]
\centering
\caption{HYM approximation measures over K3 bundle. Data obtained using commit \href{https://github.com/Justin-Tan/cymyc/commit/63bcd80417c69d224408666f051e330c08c063bb}{63bcd80}.}
\begin{tabular}{lcc}
\toprule
Quantity & Background $\nabla_0$ & Post--optimisation $\nabla$ \\
\midrule
$\langle \Lambda F \rangle$ & 
$\begin{pmatrix}
3.01 & 1.2 \times 10^{-3} \\
1.1 \times 10^{-3} & 3.00
\end{pmatrix}$ & 
$\begin{pmatrix}
3.00 & 1.3 \times 10^{-4} \\
1.7 \times 10^{-4} & 3.00
\end{pmatrix}$ \\[2ex]
\midrule
Max diagonal $\sigma$ & $0.64$ & $0.012$ \\
Max off--diagonal $\sigma$ & $0.37$ & $7.2 \times 10^{-3}$ \\
$\langle \Lambda \operatorname{Tr} F \rangle$ & $6.0 \pm 1.0$ & $6.00 \pm 0.02$\\
\bottomrule
\end{tabular}
\label{tab:hym_results_ABKO}
\end{table}

\begin{figure}
    \centering
    \includegraphics[width=1.0\linewidth]{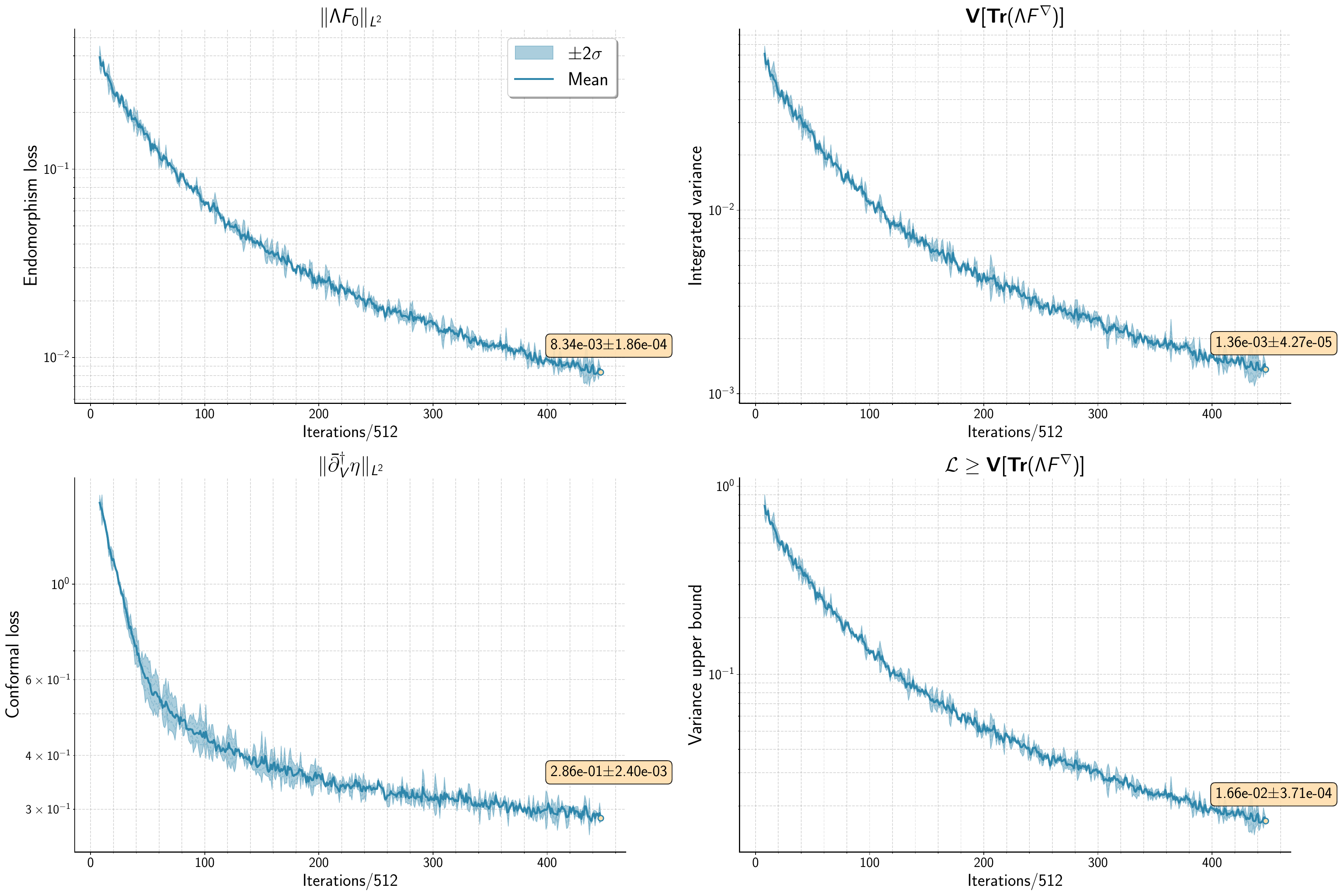}
    \caption{Evolution of various quantities during optimisation for the $\textsf{SU}(2)$ bundle over the Fermat quartic, evaluated on a separate validation set. Clockwise from top left: \textit{(a)}: Non--abelian objective function; norm of $\Lambda F_0$ \eqref{eq:ym_energy} \textit{(b)}: Integrated variance of $\Tr \Lambda F_0$ \eqref{eq:integrated_variance}, \textit{(c)}: Conformal objective function; codifferential norm \eqref{eq:codiff_11}, \textit{(d)}: Upper bound on the variance of $\Tr \Lambda F_0$ \eqref{eq:var_objective}. Results are reported on three independent runs over the same dataset.}
    \label{fig:quartic_ABKO_opt}
\end{figure}

\subsection{$\textsf{SU}(3)$ bundle over Fermat quintic}\label{sec:su3_example}
\begin{figure}
    \centering
    \includegraphics[width=1.0\linewidth]{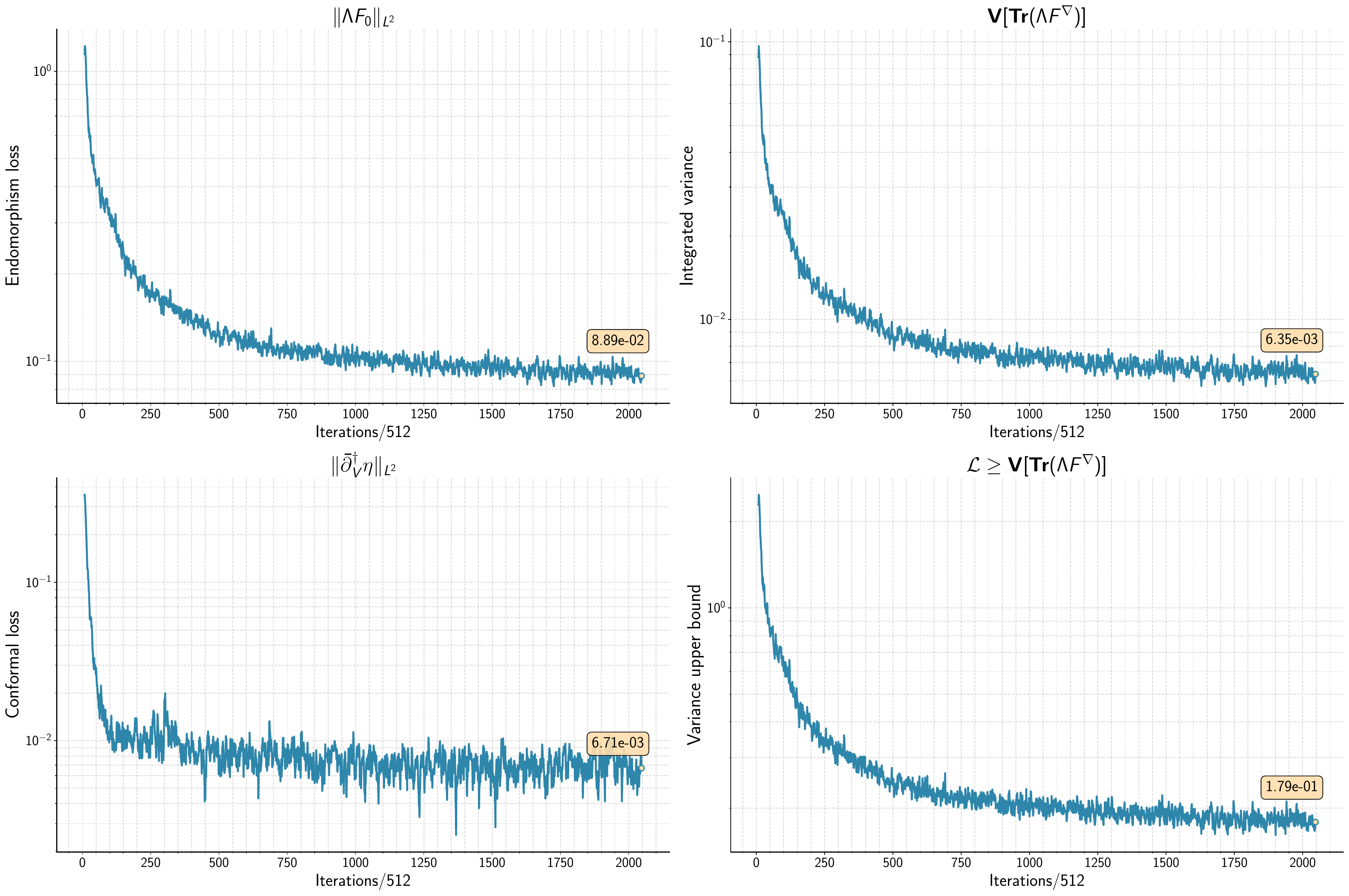}
    \caption{Evolution of various quantities during optimisation for the $\textsf{SU}(3)$ bundle over the Fermat quintic, evaluated on a separate validation set. Clockwise from top left: \textit{(a)}: Non--Abelian objective function; norm of $\Lambda F_0$ \eqref{eq:ym_energy} \textit{(b)}: Integrated variance of $\Tr \Lambda F_0$ \eqref{eq:integrated_variance}, \textit{(c)}: Conformal objective function; codifferential norm \eqref{eq:codiff_11}, \textit{(d)}: Upper bound on the variance of $\Tr \Lambda F_0$ \eqref{eq:var_objective}.}
    \label{fig:quintic_AG_opt}
\end{figure}

The last example we consider here is a $\textsf{SU}(3)$ monad bundle over the Fermat quintic \cite{Anderson:2010vdj}, defined by
\begin{equation}\label{eq:su3_quintic}
    0 \longrightarrow V \longrightarrow \bigoplus_{i=1}^4 \mathcal{O}_X(1) \overset{f}{\longrightarrow} \mathcal{O}_X(4) \longrightarrow 0~.
\end{equation}
As before, one may take the dual sequence and twist, where the bundle moduli have been fixed to $f^{\vee} = (Z_0^3, Z_1^3, Z_2^3, Z_3^3)$;
\begin{equation}
     0 \longrightarrow \mathcal{O}_X(k-4) \overset{f^t}{\longrightarrow} \mathcal{O}_X(k-1)^{\oplus 4} \longrightarrow V^{\vee} \otimes \mathcal{O}_X(k)\longrightarrow 0~.
\end{equation}
The sections are then explicitly given as the quotient
\begin{equation}
    H^0(X; V^{\vee} \otimes \mathcal{O}_X(k)) \simeq \frac{H^0(X; \mathcal{O}_X(k-1)^{\oplus 4})}{f^{\vee}_*H^0(X; \mathcal{O}_X(k-4))}~, \quad k \geq 4~. 
\end{equation}
For a bundle realised as the quotient of some ambient space $\mathcal{O}_X(k)^{\oplus K}$ via the sequence
\begin{equation*}
    0 \longrightarrow S \longrightarrow \mathcal{O}_X(k)^{\oplus K} \longrightarrow V \longrightarrow 0~,
\end{equation*}
for $X \hookrightarrow \mathbb{P}^n$, the number of global sections scales as $\dim H^0(X;V(k)) \sim K \cdot {n+k-1 \choose k} - \dim H^0(X;S \otimes \mc{L}^k)$. For this example, $\dim H^0(X;V(k)) = 139 \, (275)$ for $k=4\,(5)$, respectively. We reuse the Ricci--flat metric computed for the Fermat quintic previously, and compute the HYM connection $\nabla$ for the twisted bundle $V \otimes \mc{L}^4$. Again computing $\langle \Lambda F \rangle$ and $\mathbf{V}[\Lambda F]$ on an independent test set of $500,000$ points on $X$, we find the HYM condition to be satisfied post--optimisation within an error of $(\max \sigma_{\Lambda F^{\nabla}}) / \mu_{\Lambda F^{\nabla}} \approx 0.30\%$. The corresponding figure of merit for the background connection is $(\max \sigma_{\Lambda F^{\nabla}}) / \mu_{\Lambda F^{\nabla}} \approx 18.3\%$, and we summarise the results in Table \ref{tab:hym_results_AG}, with plots of quantities of interest depicted in Figure \ref{fig:quintic_AG_opt}.

\begin{table}[h]
\centering
\caption{HYM approximation measures over $\textsf{SU}(3)$ quintic bundle. Data obtained using commit \href{https://github.com/Justin-Tan/cymyc/commit/63bcd80417c69d224408666f051e330c08c063bb}{63bcd80}.}
\begin{adjustbox}{max width=\textwidth}
\begin{tabular}{lcc}
\toprule
Quantity & Background $\nabla_0$ & Post--optimisation $\nabla$ \\
\midrule
$\langle \Lambda F \rangle$ & 
$\begin{pmatrix}
7.50 & 7.3 \times 10^{-4} & -6.1 \times 10^{-4} \\
6.6 \times 10^{-4} & 7.50 & -4.0 \times 10^{-4} \\
-1.2 \times 10^{-3} & -7.5 \times 10^{-4} & 7.49
\end{pmatrix}$ & 
$\begin{pmatrix}
7.50 & 2.3 \times 10^{-4} & 2.4 \times 10^{-4} \\
1.2 \times 10^{-5} & 7.50 & 2.6 \times 10^{-5} \\
9.1 \times 10^{-6} & 6.3 \times 10^{-5} & 7.50
\end{pmatrix}$ \\[3ex]
\midrule
Max diagonal $\sigma$ & $1.37$ & $0.023$ \\
Max off-diagonal $\sigma$ & $0.44$ & $0.018$ \\
$\langle \Lambda \operatorname{Tr} F \rangle$ & $22.49 \pm 3.12$ & $22.50 \pm 0.011$\\
\bottomrule
\end{tabular}
\end{adjustbox}
\label{tab:hym_results_AG}
\end{table}

\section{Physical Yukawa couplings for the general embedding}\label{sec:physical_yukawa}
Here we present a computation of the physically normalised Yukawa couplings for a general embedding scenario. This is the first computation of its kind for a non--Abelian bundle. Recall the low--energy physics of a Calabi--Yau compactification is controlled by the superpotential and K\"ahler potential of the 4D action, which are entirely determined by the geometry and topology of $X$ and the associated bundle $V \rightarrow X$. The Yukawa couplings are trilinear couplings between chiral matter fields appearing in the superpotential which govern how strongly the fields interact. In particular, via the Higgs mechanism, the Yukawa couplings encode the mass spectrum of the elementary particles. One of the premises of string compactification is that all the free parameters of the Standard Model essentially have a microscopic stringy origin. After dimensional reduction of the heterotic supergravity action, the Yukawa couplings are found to be represented by a certain integral over the internal Calabi--Yau (up to an overall scaling factor) \cite{Green:1987mn}
\begin{equation}\label{eq:yukawa_couplings}
    \mathscr{Y}_{ijk} = \int_X \Omega \wedge \text{Tr}\left( \eta_i \wedge \eta_j \wedge \eta_k \right)~.
\end{equation}
This expression is quasi--topological in the sense that it only depends on the cohomology ring $H^1(X;V)$ and not the choice of representatives. Here $\text{Tr}(\cdots)$ is an instruction to form a structure group singlet from the gauge indices on $\eta$, which we have omitted. Recall that for bundles with structure group $\textsf{SU}(n)$, $\det V = \wedge^n V$ is trivial. For $\textsf{SU}(3)$ the Yukawa couplings are simply a trilinear map between matter fields in the fundamental representation (leaving the contraction over the bundle indices onto the singlet direction implicit in the first map),
\begin{equation}
    \mathscr{Y}: H^1(X;V) \otimes H^1(X;V) \otimes H^1(X;V) \rightarrow H^3(X;\wedge^3 V) \simeq H^3(X; \mathcal{O}_X) \simeq \mathbb{C}~.
\end{equation}
The physical normalisation of the Yukawa couplings is given by the matter field K\"ahler metric $\mathscr{G}_{i\overline{j}}$ appearing in the low--energy Lagrangian. Computing $\mathscr{G}_{i\overline{j}}$ requires geometric data over $X$ and $V$ through the harmonic representatives of the bundle cohomology $\tilde{\alpha} \in H^1(X;V)$. Armed with an approximation to the HYM connection on $V$, we may find the zero modes of the Dolbeault Laplacian $\tilde{\alpha}_i$ on $\Lambda^{p,q} \otimes V$, and subsequently compute 
\begin{equation}\label{eq:matter_field_metric}
    \mathscr{G}_{i \overline{j}} = \int_X \tilde{\alpha}_i \wedge \star_V \tilde{\alpha}_j~.
\end{equation}
Where the form of \eqref{eq:matter_field_metric} follows from Kaluza--Klein reduction of the heterotic supergravity theory, specifically from the expansion of the minimal coupling $A \cdot \chi \cdot \chi$ of the gaugino $\chi$ to the gauge field $A$ \cite{Green:1987mn}. To compute the proper normalisation, one must make the appropriate field redefinition such that \eqref{eq:matter_field_metric} is diagonal, $\mathscr{G} = U \Lambda U^{\dagger}$, in order to eliminate kinetic mixing terms in the low--energy Lagrangian. This amounts to finding the eigenmodes $\tilde{\Psi} = U^{\dagger} \Psi$ of $\mathscr{G}_{i\overline{j}}$, and the physically normalised Yukawa couplings are obtained as \cite{rBeast} (no sum over $i,j,k$)
\begin{equation}\label{eq:normalised_couplings}
    \tilde{\mathscr{Y}}_{ijk} := \frac{1}{\sqrt{\lambda_{i} \lambda_{j} \lambda_{k}}} U_{i}^{\phantom{i} a} U_{j}^{\phantom{j} b} U_{k}^{\phantom{k} c} \, \mathscr{Y}_{abc}~.
\end{equation}

\subsection{Couplings for an $\textsf{SU}(3)$ bundle}
We will compute a subset of the physically normalised Yukawa couplings for the $\textsf{SU}(3)$ bundle over the Fermat quintic discussed in Section \ref{sec:su3_example}. Because of the algebraic definition of such bundles, given the bundle moduli, a smooth basis of sections for any monad construction may be constructed programatically and our method readily generalises to any bundle in the monad construction. Alternatively, by appealing to the generalised Kodaira embedding theorem (Section \ref{app:embedding}), one may enumerate a basis of holomorphic sections $S \in H^0(X;V \otimes \mc{L}^k)$. Equipped with any Hermitian metric on $V(k)$, $H^{(k)} \in \Gamma(V(k)^{\vee} \otimes \overline{V(k)^{\vee}})$, one may form $S_{\flat} := H^{(k)}(\,\cdot\,, \bar{S}) \in \Gamma(V^{\vee} \otimes \mc{O}_X(-k))$, and twist by sections $P^{(k)} \in H^0(X;\mc{O}_X(k))$ to obtain a smooth basis of sections $S_{\flat} \otimes P^{(k)} \in \Gamma(V^{\vee})$.

\subsubsection{Diagram chasing}
To begin, consider the long exact sequence in cohomology associated with \eqref{eq:su3_quintic},
\begin{equation}\label{eq:su3_monad_cohomology_LES}
\begin{tikzpicture}[descr/.style={fill=white,inner sep=1.5pt}]
        \matrix (m) [
            matrix of math nodes,
            row sep=2em,
            column sep=2.5em,
            text height=1.5ex, text depth=0.25ex
        ]
        { 0 & H^0(X;V) & H^0(X; \mathcal{O}_X(1)^{\oplus 4}) & H^0(X; \mathcal{O}_X(4)) \\
            & H^1(X;V) & H^1(X; \mathcal{O}_X(1)^{\oplus 4}) & \cdots \\
        };

        \path[overlay,->, font=\scriptsize,>=latex]
        (m-1-1) edge (m-1-2)
        (m-1-2) edge node[above] {$\iota_*$} (m-1-3)
        (m-1-3) edge node[above] {$f_*$} (m-1-4)
        (m-1-4) edge[out=355,in=175,magenta] node[descr,yshift=0.3ex] {$\delta^0$} (m-2-2)
        (m-2-2) edge (m-2-3)
        (m-2-3) edge (m-2-4);
\end{tikzpicture}
\end{equation}
Notice $H^1(X; \mathcal{O}_X(1)^{\oplus 4})=0$ by Kodaira vanishing, a condition satisfied for all positive monad bundles. Hence $H^1(X; V)$ is given by the quotient of quartic polynomials by the image of linear polynomials under $f$,
\begin{equation}\label{eq:H1V_cohomology}
    H^1(X;V) \simeq \frac{H^0(X; \mathcal{O}_X(4))}{f_*\left(H^0(X; \mathcal{O}_X(1)^{\oplus 4}\right))}~.
\end{equation}
For generic choice of bundle moduli $f_i$, the map $f$ is injective and the quotient has dimension $50$. After diagram--chasing, one may write down an explicit basis $\{\xi_i\}_i$ for $H^1(X;V)$ as the projection of $\bar{\partial} \beta$, where $\beta$ denotes the preimage of the monomials in the quotient \eqref{eq:H1V_cohomology} under the monad map $f$. One may then subsequently compute the trilinear Yukawa couplings between members $\eta_i$ of the families represented by $H^1(X;V)\, , \,i=1,\ldots,\dim H^1(X;V)$. 

\subsubsection{Harmonic representatives}
Now we sketch our approximation procedure for the zero modes of the Dirac operator, which are simply harmonic bundle--valued forms on $X$. The numerical approximation of harmonic forms on $X$ taking values in a vector bundle was investigated in \cite{Berglund:2024uqv, Constantin:2024yxh} who executed this program for the standard embedding and a line bundle construction, respectively. Prior to this, semi--analytic methods for computation for $\mathbb{C}$ and line bundle--valued forms were studied in \cite{Ashmore:2020ujw, Ashmore:2023ajy}. Our method is most similar to \cite{Berglund:2024uqv}, adapted to a general embedding scenario. We approximate the harmonic representatives $\eta_i \in H^1(X;V)$ as a $\bar{\partial}_V$--exact correction to cohomologous reference representatives computed via the cohomology LES \eqref{eq:su3_monad_cohomology_LES}, which is just taken to be the basis $\{\xi_i\}$ constructed via a standard diagram chase \cite{GriffithsHarris:1994},
\begin{equation}\label{eq:form_ansatz}
    \eta_i = \xi_i + \bar{\partial}_V\mathfrak{s}_i~, \quad \mathfrak{s}_i \in \Gamma(X; V)~.
\end{equation}
Here $\mathfrak{s}_i$ is a smooth section of the bundle $V$, whose construction we defer until the end. Our objective function is simply related to the natural norm of the Dolbeault bundle Laplacian
\begin{equation}\label{eq:dolbeault_laplacian}
     \Delta_V \eta := \bar{\partial}_V^{\dagger} \bar{\partial}_V + \bar{\partial}_V \bar{\partial}_V^{\dagger}~.
\end{equation}
Notice that $\bar{\partial}_V \eta_i = 0$ by construction, so only the second term above contributes to the norm $(\eta,\bar{\partial}_V \bar{\partial}_V^{\dagger}\eta) = (\bar{\partial}_V^{\dagger} \eta,\bar{\partial}_V^{\dagger} \eta)$. We eliminate the $\eta$ scale--dependence of the Laplacian by normalising by the norm of $\eta$. In this way, we arrive quite naturally at the Rayleigh--Ritz quotient as our variational objective, parameterised by the smooth sections $\mathfrak{s}_i$;
\begin{equation}\label{eq:rayleigh_quotient}
    \mathscr{E}[\mathfrak{s}] := \frac{(\eta, \Delta_V \eta)}{(\eta, \eta)} = \frac{(\bar{\partial}_V^{\dagger} \eta,\bar{\partial}_V^{\dagger} \eta)}{(\eta, \eta)}~.
\end{equation}
Here the inner product $(\alpha, \beta) = \int_X \alpha \wedge \overline{\star}_V \beta$ is the natural one on $V$--valued forms. The outstanding question regards the parameterisation of $\mathfrak{s}$. First note that an overcomplete basis of sections $\{s_m\}_m$ of dimension $\dim H^0(X;\mathcal{O}_X(1)^{\oplus 4})$ may be programatically constructed by the bundle definition $V \simeq \ker f$, taken to be all sections in the ambient bundle $H^0(X;\mathcal{O}_X(1)^{\oplus 4})$ annihilated by the monad map $f$. $\mathfrak{s}$ is then constructed as a linear combination of such basis elements, with the coefficients $\psi$ taken to be a vector--valued smooth complex function on $X$ parameterised by the spectral network construction (Appendix \ref{app:global_fn}),
\begin{equation}
    \mathfrak{s} = \psi^m s_m \in \Gamma(X;V)~, \quad \psi \in C^{\infty}(X; \mathbb{C})^{h^1(X;V)}~.
\end{equation}
Our geometric ansatz for the zero modes of the Dirac operator now assumes the form
\begin{equation}\label{eq:harmonic_ansatz}
    \eta_i = \xi_i + \bar{\partial}_V(\psi^m_{\vartheta^*_i} s_m)~, \quad \vartheta^*_i = \argmin_{\vartheta} \mathscr{E}[(\mathfrak{s}_i)_{\vartheta}]~, \,i = 1,\ldots,\dim H^1(X;V)~.
\end{equation}
Notice this is manifestly equivariant \textit{w.r.t.} coordinate transformations for both the base $X$ and bundle $V \rightarrow X$. As we shall see, this is crucial to success of the optimisation procedure. We can efficiently vectorise the computation across the family index $i$ by using a common spectral backbone to process the input coordinates and a separate `coefficient head' $\psi_{\vartheta_i}$ to output the coefficients for each cohomology class, which obviates the need to run $\dim H^1(X;V)$ independent optimisation procedures. 

\subsubsection{Results}
\begin{figure}[htb]
    \centering
    \includegraphics[width=1.0\linewidth]{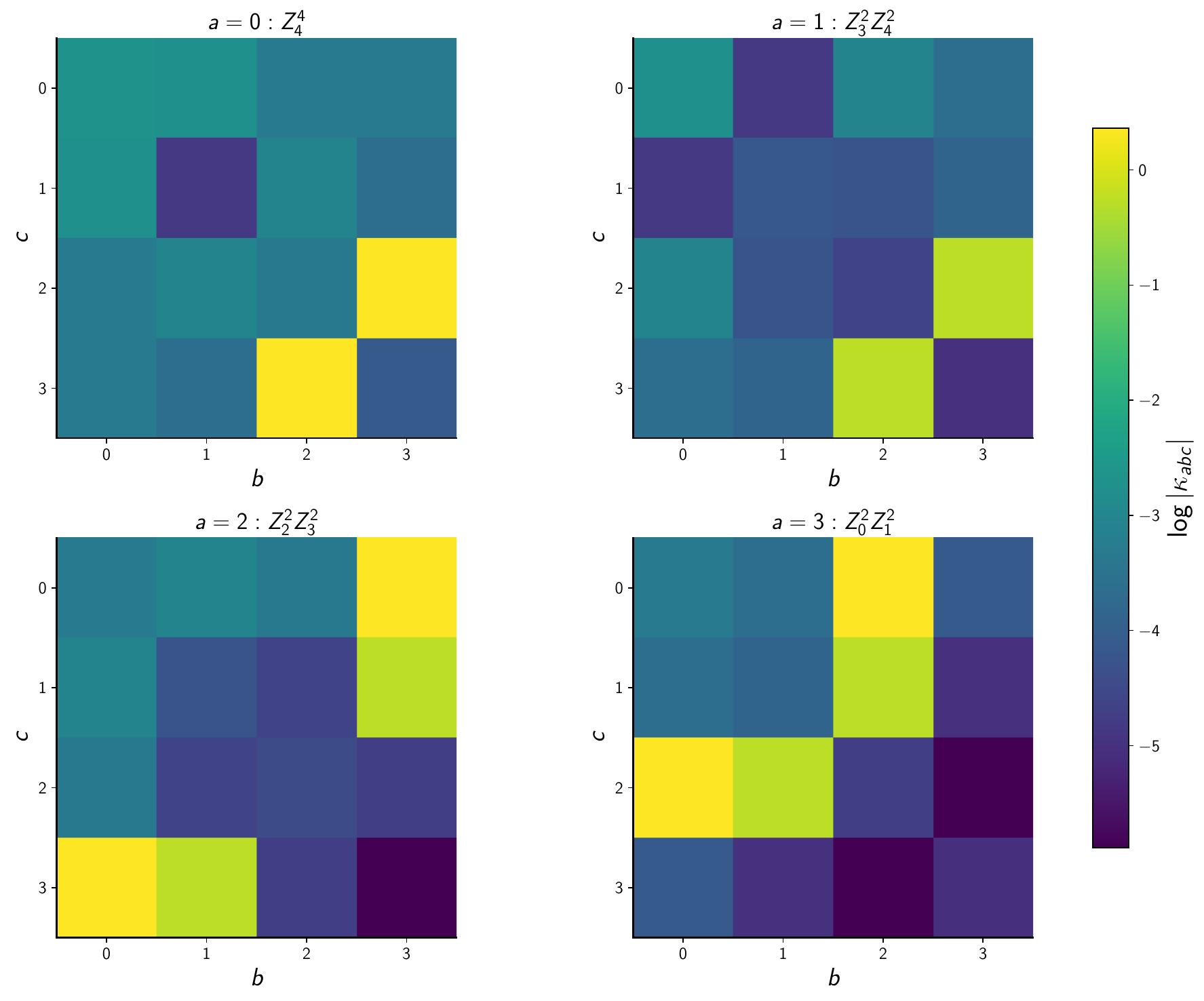}
    \caption{Heatmap of slices of the holomorphic Yukawa coupling array $\kappa_{abc}$ for four distinct classes in the cohomology $H^1(X;V)$. These classes may be represented algebraically using the indicated polynomials. This computation is semi--analytic. Note the vanishing pattern of the holomorphic couplings agrees with the pattern obtained in \cite{Anderson:2010vdj} obtained using algebraic reasoning.}
    \label{fig:holo_couplings}
\end{figure}

The bundle structure group for this example is $\textsf{SU}(3)$, leading to the visible gauge group $E_6$. In this context, the Yukawa couplings computed via \eqref{eq:yukawa_couplings} represent the four--dimensional couplings between any three given \textbf{27} $E_6$ multiplets. For numerical purposes, we fix the overall scale factor in \eqref{eq:yukawa_couplings} by noticing that $\text{Tr}\left( \eta_i \wedge \eta_j \wedge \eta_k \right) \in H^3(X;\mc{O}_X)$ is an anti--holomorphic three--form valued in the trivial bundle over $X$. This cohomology is one--dimensional and spanned by the unique $(0,3)$--form $\overline{\Omega}$ on $X$, with the Yukawa couplings simply the constant of proportionality. Thus, we have for the numerically obtained holomorphic couplings, which we denote $\kappa_{abc}$,
\begin{equation}
    \int_X \Omega \wedge \text{Tr}\left( \eta_a \wedge \eta_b \wedge \eta_c \right) = \kappa_{abc} \int_X \Omega \wedge \overline{\Omega}~.
\end{equation}
We examine four distinct classes in the cohomology \eqref{eq:H1V_cohomology}. Using the algebraic representation of these classes, they correspond to the following monomials in the polynomial basis for the quotient bundle $Q$ \eqref{eq:su3_quintic}: $\{ Z_0^4, Z_3^2Z_4^2,Z_2^2Z_3^2, Z_0^2Z_1^2\}$. First we compute the holomorphic Yukawa couplings using the reference $H^1(X;V)$ representatives computed via a diagram chase, and visualise the result in Figure \ref{fig:holo_couplings}. As a sanity check, note that the vanishing pattern computed using our differential--geometric approach agrees exactly with the pattern predicted on purely algebraic grounds in \cite{Anderson:2010vdj}.
\begin{table}[H]
\centering
\caption{Normalised Yukawa couplings \eqref{eq:normalised_couplings} $(\mu \pm \sigma)$ between eigenmodes of the matter field metric $\mathscr{G}$ \eqref{eq:matter_field_metric}, depicted as four slices through $\kappa_{abc}$. The variance is computed using jackknife resampling. Entries with significant signal--to--noise ratio $(\sigma/\mu \leq 10\%)$ are highlighted in bold. Data obtained using commit \href{https://github.com/Justin-Tan/cymyc/commit/061d1e88b994ae063274a5239579fdafe016ce5b}{061d1e8}.}
\resizebox{\textwidth}{!}{%
\begin{tabular}{ccccc|c|ccccc}
\toprule
\multicolumn{5}{c|}{\textbf{Eigenmode 1}} & & \multicolumn{5}{c}{\textbf{Eigenmode 2}} \\
\midrule
 & 1 & 2 & 3 & 4 & & & 1 & 2 & 3 & 4 \\
\midrule
1 & $0.0022 \pm 0.0027$ & $0.0021 \pm 0.0017$ & $0.080 \pm 0.043$ & $0.103 \pm 0.057$ & &
1 & $0.0021 \pm 0.0017$ & $0.0038 \pm 0.0017$ & $\mathbf{0.205 \pm 0.011}$ & $\mathbf{0.277 \pm 0.014}$ \\
2 & $0.0021 \pm 0.0017$ & $0.0038 \pm 0.0017$ & $\mathbf{0.205 \pm 0.011}$ & $\mathbf{0.277 \pm 0.014}$ & &
2 & $0.0038 \pm 0.0017$ & $0.0057 \pm 0.0028$ & $0.077 \pm 0.042$ & $0.099 \pm 0.057$ \\
3 & $0.080 \pm 0.043$ & $\mathbf{0.205 \pm 0.011}$ & $0.0035 \pm 0.0021$ & $0.0038 \pm 0.0022$ & &
3 & $\mathbf{0.205 \pm 0.011}$ & $0.077 \pm 0.042$ & $0.0039 \pm 0.0028$ & $0.0081 \pm 0.0024$ \\
4 & $0.103 \pm 0.057$ & $\mathbf{0.277 \pm 0.014}$ & $0.0038 \pm 0.0022$ & $0.0037 \pm 0.0025$ & &
4 & $\mathbf{0.277 \pm 0.014}$ & $0.099 \pm 0.057$ & $0.0081 \pm 0.0024$ & $0.0046 \pm 0.0025$ \\
\midrule
\midrule
\multicolumn{5}{c|}{\textbf{Eigenmode 3}} & & \multicolumn{5}{c}{\textbf{Eigenmode 4}} \\
\midrule
 & 1 & 2 & 3 & 4 & & & 1 & 2 & 3 & 4 \\
\midrule
1 & $0.080 \pm 0.043$ & $\mathbf{0.205 \pm 0.011}$ & $0.0035 \pm 0.0021$ & $0.0038 \pm 0.0022$ & &
1 & $0.103 \pm 0.057$ & $\mathbf{0.277 \pm 0.014}$ & $0.0038 \pm 0.0022$ & $0.0037 \pm 0.0025$ \\
2 & $\mathbf{0.205 \pm 0.011}$ & $0.077 \pm 0.042$ & $0.0039 \pm 0.0028$ & $0.0081 \pm 0.0024$ & &
2 & $\mathbf{0.277 \pm 0.014}$ & $0.099 \pm 0.057$ & $0.0081 \pm 0.0024$ & $0.0046 \pm 0.0025$ \\
3 & $0.0035 \pm 0.0021$ & $0.0039 \pm 0.0028$ & $0.0028 \pm 0.0038$ & $0.0010 \pm 0.0034$ & &
3 & $0.0038 \pm 0.0022$ & $0.0081 \pm 0.0024$ & $0.0010 \pm 0.0034$ & $0.0061 \pm 0.0033$ \\
4 & $0.0038 \pm 0.0022$ & $0.0081 \pm 0.0024$ & $0.0010 \pm 0.0034$ & $0.0061 \pm 0.0033$ & &
4 & $0.0037 \pm 0.0025$ & $0.0046 \pm 0.0025$ & $0.0061 \pm 0.0033$ & $0.0048 \pm 0.0038$ \\
\bottomrule
\end{tabular}
}
\label{tab:four_classes}
\end{table}

\begin{figure}[htb]
    \centering
    \includegraphics[width=1.0\linewidth]{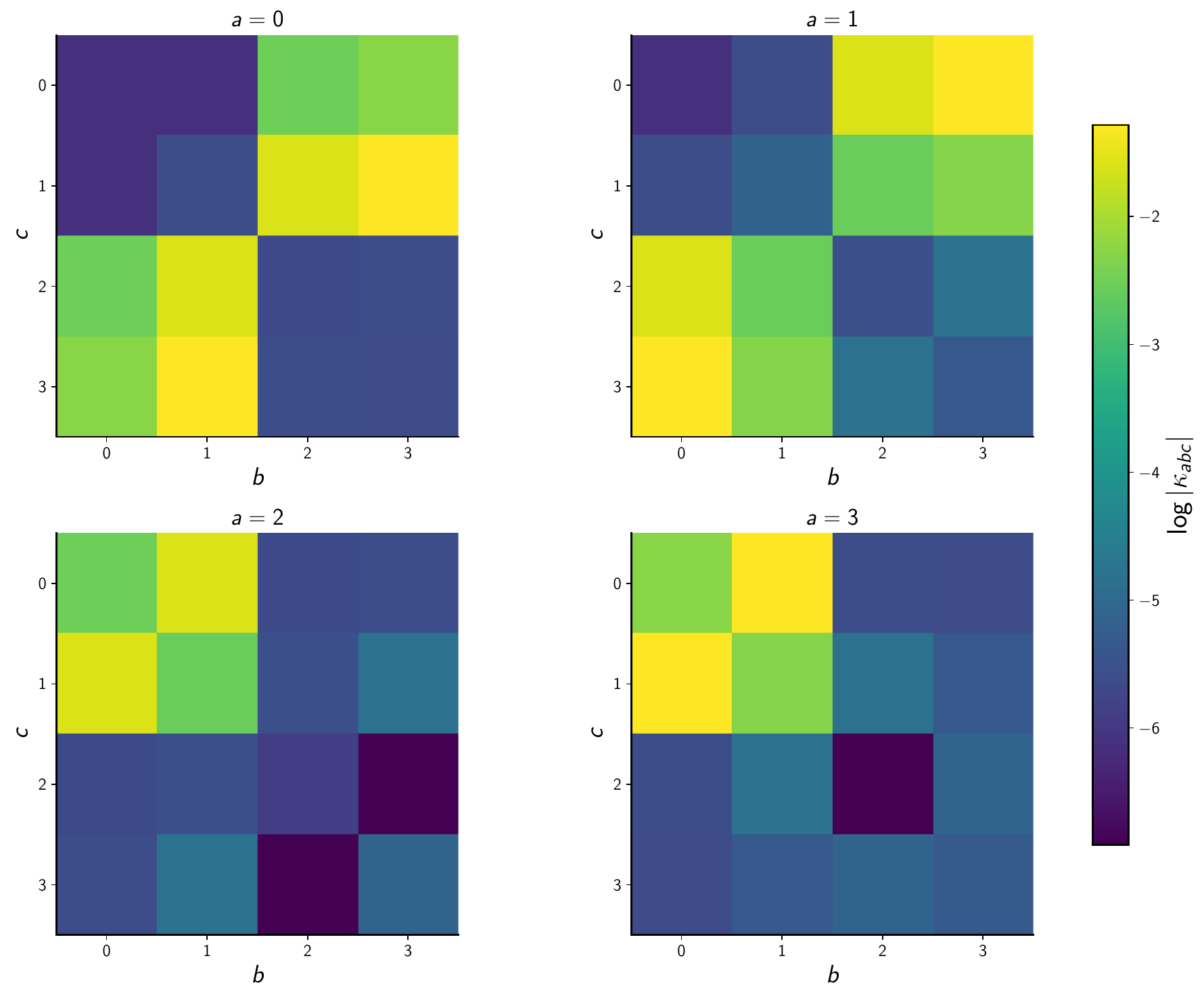}
    \caption{Heatmap of slices of the normalised Yukawa coupling array $\widetilde{\kappa}_{abc}$, evaluated in the eigenbasis of the matter field metric $\mathscr{G}_{a\overline{b}}$ \eqref{eq:matter_field_metric}. This metric is evaluated using four distinct classes in the cohomology $H^1(X;V)$. This computation is differential--geometric and depends on our learned approximations.}
    \label{fig:norm_couplings}
\end{figure}

Using the approximate harmonic forms $\tilde{\alpha}$ obtained from our procedure \eqref{eq:harmonic_ansatz}, we compute the matter field metric $\mathscr{G}_{a\overline{b}}$ \eqref{eq:matter_field_metric} via Monte Carlo integration over $X$ using an independent sample of $2.5 \times 10^6$ points. We rotate to the eigenbasis of $\mathscr{G}$ and compute the normalised couplings \eqref{eq:normalised_couplings}, exhibited in Table \ref{tab:four_classes} and visualised in Figure \ref{fig:norm_couplings}. Keeping in line with the numerical observations of \cite{Butbaia:2024tje, Berglund:2024uqv}, the physical normalisation appears to have a qualitatively `polarising' effect in that the canonical normalisation promotes certain combinations of couplings while heavily suppressing others. We include further visualisations of the couplings for this particular subset of four families, as well as a larger set of 9, 16 and 25 families in Appendix \ref{app:coupling_vis}. 

To emphasise the importance of equivariance \eqref{eq:equivariant_nn}, we compare our geometric ansatz \eqref{eq:harmonic_ansatz} to one which is not equivariant by construction but instead encourages global coherency through an auxiliary transition loss in Appendix \ref{app:global_versus_local}. 

The evolution of quantities of interest during optimisation for the harmonic representatives are exhibited in Appendix \ref{app:harmonic_opt_plots}. In particular, we note that the eigenvalues of the K\"ahler matter field metric $\mathscr{G}$ rapidly converge to their steady--state values during optimisation and exhibit near--negligible variance, as exhibited in Figures \ref{fig:G_eigs_4} and \ref{fig:G_eigs_16}. Recall the eigenbasis of $\mathscr{G}$ is exactly the salient information to be derived from this procedure, to be used to compute the physically normalised couplings \eqref{eq:normalised_couplings}. This suggests that obtaining a reliable approximation to physical observables may not necessitate onerous computation for each point in moduli space, which may enable us to efficiently scan through regions of the combined moduli space in future large--scale studies. 

\section{Conclusions and future work}
Our goal was to solve the Hermitian Yang--Mills equations \eqref{eq:hym} on a holomorphic vector bundle $V \rightarrow X$. This is a challenging nonlinear elliptic system, but we have shown that elementary differential--geometric considerations can be rather decisive on this matter. In particular, embedding the geometric constraints directly into the neural ansatz used to model the connection provides an effective means of solving this system. We have demonstrated an alternating optimisation procedure which numerically approximates the HYM connection on stable holomorphic bundles $V$ to within $\mc{O}(0.1\%)$ precision. At each stage, we construct an appropriate equivariant neural ansatz for the required geometric data via the Kodaira embedding of $X$ into some suitable Grassmannian determined by the bundle $V$. Our approach applies to any holomorphic bundle that admits a HYM connection provided one may explicitly write down this embedding. We have illustrated our approach on a general class of vector bundles obtained via the monad construction. Our work has a clear phenomenological application to the study of string compactifications. We now find ourselves armed with the ability to forge a connection between the data of the compactification and quantitative physical observables for string models involving a non--Abelian gauge group, and hope this motivates further data--driven exploration of the vast string landscape. Beyond direct phenomenology, it would be of interest to apply these numerical techniques to investigate the structure of the matter field K\"ahler potential and superpotential couplings of heterotic theories \cite{Anderson:2021unr, Gray:2025dao}, in the hope of revealing vanishing patterns that may be studied analytically. In general the normalised couplings \eqref{eq:normalised_couplings} are a complicated function of the bundle, complex structure and K\"ahler moduli, and studying the moduli dependence of $\kappa_{abc}$ may uncover interesting structure within the combined moduli spaces.

The main drawback of our approach is the polynomial growth of $N_k = \dim H^0(X;V \otimes \mc{L}^k)$ in $k$. The computational time and memory requirements of the forward pass through the neural network ansatz for the non--Abelian stage \eqref{eq:h_ansatz} scales as $\mc{O}(N_k^2) \sim \mc{O}(k^{2n})$. As computing the curvature two--form involves the Hessian $\bar{\partial}\partial$ of the endomorphism network \eqref{eq:curvature_correction}, our algorithm is prohibitively expensive for bundles which require a large degree of twisting such that $V \otimes \mc{L}^k$ is globally generated. While reasonably performant for a given choice of moduli for the bundle $V \rightarrow X$, this precludes the application of our procedure at scale to large classes of phenomenologically interesting bundles across a range of bundle and complex structure moduli. Arguing by analogy with an expansion in Fourier modes, it is unlikely that all the basis sections of $V \otimes \mc{L}^k$ contribute equally to the hypothesis \eqref{eq:h_ansatz}. A modest reduction, linear in $k$, in the number of sections we must consider to attain similar performance would translate to a quadratic decrease in space complexity. Bearing this in mind, it would be fruitful to identify and work directly with an effective subspace of sections of $V \otimes \mc{L}^k$ by extending the optimisation procedure to the associated Grassmannian \cite{Ek:2024fgd}. Lastly, we will use the methods illustrated within to derive the physically normalised Yukawa couplings for a collection of phenomenologically interesting string compactification scenarios in forthcoming work.

\section*{Acknowledgements}
The authors would like to thank Anthony Ashmore, Per Berglund, Giorgi Butbaia,  Yang--Hui He, Tristan H\"ubsch, Vishnu Jejjala, Oisin Kim and Dami\'an Mayorga Pe\~na for helpful discussions. They would also like to thank the organisers of the Pollica workshop on Calabi--Yau manifolds, where the idea for this work originated. JT acknowledges the support of the Accelerate Programme for Scientific Discovery. 

\bibliographystyle{utphys} 
\bibliography{ref}

\newpage

\begin{appendices}

\section{Conventions}\label{app:conventions}
The setting is a rank-$d$ holomorphic vector bundle over an $n$--dimensional K\"ahler manifold. $V \rightarrow (X,\omega)$. $X \subset \mathbb{P}^{n_1} \times \cdots \times \mathbb{P}^{n_K}$ will be taken to be a Calabi--Yau manifold realised as a complete intersection. That is, embedded in some ambient product of projective space factors by a system of homogeneous polynomials of the requisite degree. Latin, Greek indices denote bundle and base indices respectively. We may also use $(i,j,k,l)$ for base indices --- this will be clear from context. We will adopt the convention that the first bundle index will always denote the row index and the second denotes the column index for the matrix representation of a tensor field; \textit{e.g.} for the connection we have locally $A_{a\phantom{b}\mu}^{\phantom{a} b}$ and for the curvature $F_{a \phantom{b} \mu \overline{\nu}}^{\phantom{a} b}$. To disambiguate between curvature forms arising from different choices of connection $\nabla$ on $V$ we will use a superscript $F^{\nabla}$ where necessary. We will always work with the Chern connection with respect to the Hermitian structure $H$ on $V$. This leads to our convention for the curvature as
\[ F = \bar{\partial}(\partial H H^{-1})~, \]
to be compared with the traditional physics convention $F = \bar{\partial}(H^{-1}\partial H)$. For numerical computations we adopt the following conventions for the degree $d(V,\omega)$ and slope $\mu(V, \omega)$ (but we elide factors of $2\pi i$ in the body of the text);
\begin{equation}\label{eq:slope_conventions}
    \Lambda F = 2 \pi i \mu \mathbf{1}_V~, \quad d(V, \omega) = \int_X c_1(V) \wedge \omega^{n-1} = \frac{i}{2\pi}\int_X \Tr F \wedge \omega^{n-1}~.
\end{equation}
Here the dual Lefschetz operator $\Lambda$ on $(1,1)$ forms is the contraction w.r.t. K\"ahler form $\omega$. By working in K\"ahler normal coordinates, $\Lambda F = (F \wedge \omega^{n-1}) / \omega^n$. Integrating the first identity, the slope in our conventions is
\begin{equation}
    \mu(V,\omega) = - \frac{1}{n!}\frac{d(V,\omega)}{\rank V \cdot \text{Vol}(X)}~.
\end{equation}

\section{Kodaira embedding}\label{app:embedding}
One natural parameterisation of the Hermitian metric on $V$ is via an expansion in a basis of global sections of $V$. However, a stable bundle $V$ with vanishing first Chern class is absent global holomorphic sections; $H^0(X;V) = 0$. The gauge bundles considered in heterotic string compactifications have structure group $SU(n)$ and hence $c_1(V) =0$ by definition. By the Kodaira embedding theorem, one may instead enumerate a basis of global sections for $V$ tensored with a sufficiently larger power of an ample line bundle $\mc{L}$. Denote this as $V \otimes \mc{L}^k =: V(k)$. The generalised Kodaira embedding theorem \cite{Ma2007} states that, for a compact complex manifold $X$ equipped with a holomorphic vector bundle $V$, for $k$ sufficiently large, a basis $\{S_m\}$ of the space $H^0(X; V \otimes \mc{L}^k)$ induces the embedding $\iota_k: X \hookrightarrow Gr(r, N_k)$, 
\begin{equation}
\label{eq:keller_diagram}
\begin{tikzcd}
	{V \otimes \mc{L}^k} \arrow[r, "\Phi"] \arrow[d] & {\mathbf{U}_{r,N_k}} \arrow[d] \\
	X \arrow[r, hookrightarrow, "\iota_k"] & Gr(r, N_k)~.
\end{tikzcd}
\end{equation}
Here $N_k = \dim H^0(X; V \otimes \mc{L}^k)$, and $U_{r,N_k}$ is the dual of the universal bundle over the Grassmannian. To be computationally explicit, $\iota_k$ defines a map from $X$ into $H^0(X; V \otimes \mc{L}^k)$, regarded as a complex vector space;
\begin{equation}\label{eq:k_embedding}
    x \longmapsto \{\mathbf{S}^a_i\} :=\left[ \begin{pmatrix} S_1^1(x) \\ \vdots \\ S_1^r(x) \end{pmatrix} : \cdots :\begin{pmatrix} S_{N_k}^1(x) \\ \vdots \\ S_{N_k}^r(x) \end{pmatrix} \right]
\end{equation}
Here $a = 1, \ldots, r$ is the colour index for $V$ and $i$ indexes the number of global sections of $V \otimes \mc{L}^k$. Given a choice of local frame, \eqref{eq:k_embedding} is locally described by an $r \times N_k$ matrix, and assigns to each $x \in X$ an $r$--dimensional subspace of $H^0(X; V \otimes \mc{L}^k)$, defined up to a $\textsf{GL}(r)$ choice of frame, which defines a point in $Gr(r, N_k)$. In addition, one may also make a $\textsf{GL}(N_k)$ choice of basis for $H^0(X; V \otimes \mc{L}^k)$ --- generically changing the point in $Gr(r,N_k)$. This will become important for numerical stability, as we will see in the sequel. The embedding \eqref{eq:keller_diagram} furnishes us with a basis of global sections needed to construct tensor fields over $X$ taking values in the twisted bundle $V(k)$. 

Now a positive--definite Hermitian form on the dual vector space of global sections, $H^0(X;V(k))^{\vee}$, defines a Hermitian metric on the universal bundle over the Grassmannian. This is a generalisation of the Fubini--Study metric induced by the Kodaira embedding into $\mathbb{P}^n$ when $V$ is a line bundle. From now we will work with $H^0(X;V(k))$ directly instead of $H^0(X;V(k))^{\vee}$, with the understanding we may always translate between the two. Concretely, one may fix a reference Hermitian metric $H_V$ on $V$ and $h_{\mc{L}}$ on $\mc{L}$. This induces a reference Hermitian metric $H_V \otimes h_{\mc{L}}^{\otimes k}$ on $V(k)$, which in turn defines a reference Hermitian form on $H^0(X;V(k))$ via the associated $L_2$ inner product, up to some normalisation factor,
\begin{equation}\label{eq:H_form}
    \mathbf{H}_{\textsf{ref}}(s_a, s_b) := \langle s_b \vert s_a\rangle_{H^0(V(k))} \sim \int_X \text{dVol}_X \, \langle s_b\vert s_a\rangle_{V(k)}~.
\end{equation}
One may associate a Hermitian metric $\textsf{FS}(\mathbf{H})$ on $V(k)$ to the Hermitian form $\mathbf{H}_{\textsf{ref}}$ on $H^0(V(k))$ --- this will be in general different from the metric induced by $H_V$. The full details of the construction may be found in \cite{wang2005, keller:2006}, and we sketch this briefly below. First note for any point $p \in X$, there is a natural evaluation map which takes a global section $S \in H^0(X;V(k))$ and returns its value at $p$,
\begin{align*}
    \textsf{ev}:& \, H^0(X;V(k)) \times X\rightarrow V(k) \vert_p\\
       &(S,p) \mapsto S(p)
\end{align*}
Define $Q$ as the adjoint of $\textsf{ev}(\cdot,p)$ w.r.t. $H_{\textsf{ref}}$, sending a vector in the fibre to a section,
\begin{equation*}
    Q: V(k)\vert_p \rightarrow H^0(X;V(k))~.
\end{equation*}
One may then define a Hermitian metric on $V(k)$ as
\begin{equation}\label{eq:fs_ref}
    h_k :=Q^{\dagger} Q \in \Gamma\left(X; V(k)^{\vee} \otimes V(k)^{\vee}\right)~, \quad h_k(S,S') = \left\langle Qs, Qt\right\rangle_{\mathbf{H}_{\textsf{ref}}}~, \; s,t \in V(k)\vert_p~.
\end{equation}
$h_k$ agrees with the pullback by $\Phi$ of the Fubini--Study metric on the universal bundle over $Gr(r, N_k)$, defined by the Hermitian form $\mathbf{H}_{\textsf{ref}}$ on $H^0(X;V(k))$. We refer to this as the reference Fubini--Study metric defined by $\mathbf{H}_{\textsf{ref}}$. Note that choosing $\mathbf{H}_{\textsf{ref}}$ to be the identity on $H^0(X;V(k))$ yields a canonical reference Fubini--Study metric on the twisted bundle $V(k)$ which will be of use in the following. Any pair of Hermitian forms on the vector space $H^0(V(k))$ are related by the action of some $\sigma \in \textsf{GL}\left(H^0(V(k))\right)$, and each of these forms defines a unique Hermitian metric on $V(k)$. In similar notation to \eqref{eq:fs_ref},
\begin{equation}
    h_k(\sigma) :=Q^{\dagger}\sigma^{\dagger} \sigma Q \in \Gamma\left(X; V(k)^{\vee} \otimes V(k)^{\vee}\right)~,
\end{equation}
which is just the Fubini--Study metric on $V(k)$ defined by the form $\sigma^{\dagger} H \sigma$. This gives a correspondence between Hermitian metrics on $V(k)$ and Hermitian forms on the space of global sections $H^0(X;V(k))$. The utility of this construction comes from the asymptotic Tian--Yau--Zelditch expansion \cite{Ma2007}. Let $\mc{H}_{\infty}$ denote the space of smooth Hermitian metrics on $V(k)$, and $\mc{H}_k$ denote the subset of $\mc{H}_{\infty}$ consisting of Fubini--Study metrics defined by a Hermitian form on $H^0(X;V(k))$. The Tian--Yau--Zelditch theorem states that the union of all $\mc{H}_k$ for $k$ sufficiently large is dense in $\mc{H}_{\infty}$,
\begin{equation}\label{eq:tyz}
    \mc{H}_{\infty} = \overline{\cup_{k \gg 0} \mc{H}_k}~.
\end{equation}
In other words, given some $p\in \mathbb{N}$, any element of $h \in \mc{H}_{\infty}$ may be approximated arbitrarily well by a sequence of Fubini--Study metrics $\{h_{k,p}\}_{k \in \mathbb{N}}, \,h_{k,p} \in \mc{H}_k$ by choosing $k$ sufficiently large; $h_{k,p} \rightarrow h$ as $k \rightarrow \infty$ in the $C^p$--norm. This may be used to simplify the problem of numerical approximation of Hermite--Einstein metrics on a stable vector bundle \eqref{eq:hym} \cite{Douglas:2006hz, Anderson:2010ke} as it reduces the search space for such metrics from the infinite--dimensional space $\mc{H}_{\infty}$, to the finite--dimensional space $\mc{H}_k$ parameterised by constant Hermitian matrices on $X$ \eqref{eq:H_form} via the map $\textsf{FS}$.

\section{Global function parameterisation}\label{app:global_fn}
When undertaking numerics on a manifold $X$, it is important to distinguish between global and local data. One unavoidably has to choose a coordinate chart for $X$ to perform numerical computations. Unlike local information, global data is independent of this arbitrary choice of chart and may be used to define intrinsic, physically meaningful geometric quantities. For these reasons, we will often need to parameterise functions which are globally defined over the entire manifold. This may be efficiently achieved by an appropriate basis expansion of the input \cite{Berglund:2022gvm}, as we review below. We are interested in the case where $X$ is a K\"ahler manifold defined as the zero locus of generic homogeneous polynomials $\{f_i\}_{i=1,\dots,N}$ where $f_i \in \IC[Z_0,\dots,Z_{n_i}]$, thus $X$ lies in $\mathbb{P}_\IC^{n_1}\times \dots \times \mathbb{P}_\IC^{n_N}$. For each component $\mathbb{P}_\IC^{n_i}$ of the product, define a mapping:
\begin{gather}
	\alpha_{n_i}\colon \mathbb{P}_\IC^{n_i}	\longrightarrow \mathbb{C}^{n_i+1,n_i+1} ~,
\end{gather}
whose action on a general point $p\in [Z_0\colon Z_1\colon \cdots \colon Z_{n_i}]\in\mathbb{P}_\IC^{n_i}$ is defined as:
\begin{gather}\label{eq:spec_embed}
	\alpha_{n_i}(p) = \left[\begin{matrix}
		\displaystyle \frac{Z_0 \overline{Z_0}}{|Z|^2} && \displaystyle\frac{Z_0 \overline{Z_1}}{|Z|^2} && \dots && \displaystyle\frac{Z_0 \overline{Z_{n_i}}}{|Z|^2}  \\
		\displaystyle\frac{Z_1 \overline{Z_0}}{|Z|^2}  && \displaystyle\frac{Z_1 \overline{Z_1}}{|Z|^2} && \dots && \displaystyle\frac{Z_1 \overline{Z_{n_i}}}{|Z|^2} \\
		\vdots && \vdots && \ddots && \vdots \\
		\displaystyle \frac{Z_{n_i} \overline{Z_0}}{|Z|^2} && \displaystyle\frac{Z_{n_i} \overline{Z_1}}{|Z|^2} && \dots && \displaystyle\frac{Z_{n_i} \overline{Z_{n_i}}}{|Z|^2}
	\end{matrix}\right] ~.
\end{gather}
Note that $\alpha_{n_i}$ is a well--defined global smooth function on $\mathbb{P}_\IC^{n_i}$ and thus its restriction $\alpha_{n_i}\vert_{X}$ is a well-defined smooth function on $X$. Furthermore, note that the components of $\alpha_{n_i}$ correspond to the second--order basis of monomials used in~\cite{PhysRevD.103.106028} to build the eigenfunctions of the Laplace operator $\Delta$. Thus we shall refer to the expansion \eqref{eq:spec_embed} as a spectral embedding and the corresponding neural network as a spectral neural network.

\section{Experimental details}\label{app:experiments}
All experiments were run on a single GPU machine equipped with an NVIDIA RTX A5000 card with 24GB of VRAM and an Intel Xeon Gold 5220R processor with 32GB of RAM. All quantitative results involving Monte Carlo integration are reported on an independent test set consisting of 250,000 points sampled from the relevant manifold, unless otherwise stated. The code for our experiments is open--source, written predominantly in \texttt{Jax} \cite{jax2018github} for efficient numerical evaluation of derivatives using automatic differentiation, and may be accessed at \url{https://github.com/Justin-Tan/cymyc}.

\subsection{Ricci--flat representative optimisation}
\subsubsection{Fermat quintic}
For bundles over the Fermat quintic, defined in terms of the hypersurface equation
\begin{equation}
    \left\{\sum_{i=0}^4 Z_i^5 - 5 \psi \prod_i Z_i = 0 \right\} \subset \mathbb{P}^4~,
\end{equation}
we always work at the moduli point $\psi=0$. We optimise $\varphi_{\textsf{NN}}: X \rightarrow \mathbb{R}$ for 256 epochs on a training set of $2 \times 10^6$ points sampled from $X$. The architecture for our model consists of a `spectral embedding' stage ---  an expansion of the input homogeneous coordinates in the degree-2 elements of the basis of eigenfunctions on $X$ \eqref{eq:spec_embed}. This is followed by 3 fully connected layers of widths $(42,42,42)$ using the GeLU activation function. Unless otherwise stated, the network parameters for all models will be optimised using the \textsf{AdamW} optimiser with a learning rate of $10^{-4}$. We repeat this experiment independently three times, and obtain a sigma--measure of $\sigma = (1.2 \pm 0.2) \times 10^{-3}$.

\subsubsection{Fermat quartic}
For the Fermat quartic, defined in terms of the hypersurface
\begin{equation}
    \left\{\sum_{i=0}^3 Z_i^4 - 4 \psi \prod_i Z_i = 0 \right\} \subset \mathbb{P}^3~,
\end{equation}
we work at the moduli point $\psi = 0.5 e^{2\pi im/4}$, using the same dataset size, architecture and optimisation hyperparameters in the Fermat quintic case, obtaining a sigma--measure of $\sigma = (1.4 \pm 0.1) \times 10^{-3}$.

\subsection{HYM connection approximation}
Recall that we divide the procedure of finding the HYM connection into two stages (Section \ref{sec:objective}). 
\begin{itemize}
    \item The first `Abelian' stage corresponds to prescribing the curvature on $\det V$ s.t. $\Tr F^{\nabla}$ is harmonic. This can always be achieved on a compact K\"ahler manifold through a conformal transformation $H \mapsto e^fH$, where the global function $f \in C^{\infty}(X)$ is parameterised as a spectral neural network consisting of the spectral embedding \eqref{eq:spec_embed} followed by 3 fully connected layers of widths $(48,48,48)$ using the GeLU activation function.
    \item The second `non--Abelian' stage outputs a Hermitian matrix representing the coefficients of an expansion $H^{mn}, m,n=1,\ldots,\dim H^0(X; V(k))$ in a basis of sections for the endomorphism section \eqref{eq:h_ansatz}. We construct $H^{mn}$ via the square--root free Cholesky decomposition $H = LDL^{\dagger}$. The complex lower triangular matrix $L$ and real diagonal vector $D$ are both parameterised using an spectral embedding followed by three fully connected layers of widths $(48,48,48)$. The final layer outputs $N_k^2$ complex numbers representing the number of degrees of freedom in the Hermitian coefficient matrix in the ansatz \eqref{eq:h_ansatz}.
\end{itemize}

\subsubsection{Tensor decomposition}
For bundles with $\dim H^0(X; V(k)) \gg 1$, computing the curvature tensor \eqref{eq:curvature_correction} induced by the endomorphism may become prohibitively expensive as the implementation involves nested Jacobian calls. An alternative to parameterising the full coefficient matrix with $N_k \times N_k$ degrees of freedom is to resort to a low--rank approximation for the coefficient matrix, and parameterise instead a complex matrix $K \in \mathbb{C}^{N_k \times q}, \, q \ll N_k$ and a real diagonal such that $H = K K^{\dagger} + D$. This decreases the size of the final layer parameterising the Hermitian matrix from $N_k^2 \rightarrow N_k \times q$. Empirically, we find decreased convergence speed but little to no difference in final performance with the full uncompressed matrix for $q \sim (K/8, \ldots, K/2)$. This perhaps indicates that only a minority of the elements of a generic basis of sections for $V(k)$ significantly contribute in forming the hypothesis \eqref{eq:h_ansatz}. It may be useful to consider models with sparse connectivity patterns, or which implement some weight sharing mechanism to reduce the memory requirements of the nested Jacobian operations required to instantiate the curvature tensor. 

\subsubsection{Alternate objective functionals}
For the Abelian optimisation stage, we obtain qualitatively similar results using the following objective functional, proposed in \cite{Ashmore:2021rlc}.
\begin{equation}\label{eq:var_objective}
    \mathcal{L}[h,\omega] = \left\langle \Tr \left(\Lambda F_0\right)^2 \right\rangle - \frac{1}{\rank V} \left\langle \Tr \Lambda F_0\right\rangle^2~.
\end{equation}
Where the expectation $\langle \cdots \rangle$ is computed w.r.t. to the determinant volume form $\omega^n / n!$ induced by the K\"ahler metric. Owing to the Cauchy--Schwarz inequality for a diagonalisable matrix $A$, $d \cdot \Tr A^2 \geq (\Tr A)^2$, one has
\[ \mathcal{L}(h, \omega) \geq \frac{1}{\rank V} \mathbf{V}\left[\Tr \Lambda F_0\right] \geq 0~.\]
Optimising \eqref{eq:var_objective} is then equivalent to minimising an upper bound on the variance of the trace $\Tr \Lambda F_0 \in C^{\infty}(X; \mathbb{R})$, and vanishes iff $\Lambda F_0$ is the zero matrix. In principle any objective which encourages the vanishing of some norm of $\Lambda F_0$, together with constancy of the trace $\Lambda \Tr F$ will be viable here. Empirically, we found the codifferential objective more effective at reducing $\mathbf{V}[\Lambda \eta]$ \eqref{eq:integrated_variance}, despite not directly optimising for the variance of the trace. 
\subsubsection{Complexity considerations}
Computing the codifferential is significantly more expensive as gradient optimisation of this objective is third--order in $f$. As we have to instantiate the entire Jacobian in the computation of \eqref{eq:codiff_11}, the memory complexity scales as $O(m^3 \times W)$ (as the $\ell$--th nested Jacobian is built out of $m^{\ell}$ Jacobian--vector--product calls), where $m = \dim_{\mathbb{R}} X$ is the input dimension and $W$ is the maximum width of the neural networks used to model $(f,H)$. This scales polynomially with the twisting degree $k$ --- the maximal width $W$ in general will correspond to the degrees of freedom used to model the Hermitian matrix in \eqref{eq:h_ansatz}. This scales as $N_k^2$, where $N_k \sim O(k^n)$. Thus, memory--wise, using the codifferential loss \eqref{eq:codiff_11} will be infeasible for large values of $k$, and \eqref{eq:integrated_variance} may be used as a substitute. 

Regarding the non--Abelian stage, examining the gradient flow of the Yang--Mills energy \eqref{eq:ym_energy}, the corresponding evolution equation of is of order four in the bundle metric $H$. It would be computationally advantageous to develop a static functional for this stage with a lower--order evolution equation.

\subsection{Harmonic representative optimisation}
We use a simple densely connected architecture of $(48,64,48)$ units as the spectral network backbone, while each coefficient head indexed to the relevant cohomology class projects the final layer of the spectral network to a $\dim E$--dimensional representation. We use a learning rate of $2.5 \times 10^{-4}$ in all experiments with the \texttt{AdamW} optimiser. 


\section{Untwisting}\label{app:untwisting}
The output of our procedure is the HYM connection on the twisted bundle $V \otimes \mathcal{L}^k$. We would like to recover the Hermitian structure associatd with the Chern connection, $H_V$, on the original bundle $V$, which is related to the twisted bundle metric by
\begin{equation}
    H_{V \otimes \mathcal{L}^k} = H_V \times H_{\mathcal{L}}^k~.
\end{equation}
Recall the curvature of a twisted bundle only picks up a contribution to the trace,
\[ F_{V \otimes \mathcal{L}^k} = F_V + k F_{\mathcal{L}} \cdot \mathbf{1}_V~. \]
To obtain the connection on $V$, one must subtract the Abelian contribution from $\mathcal{L}^k$. As discussed in \cite{Anderson:2010ke}, there is freedom in the choice of fiber metric on $\mathcal{L}$ used for this. Fortunately, given the Hermitian structure on the twisted bundle, the untwisting procedure is entirely self--contained; using the fiber metric $H_{\mathcal{L}} = (\det H_{V \otimes \mathcal{L}^k})^{1/m}$ induced by the determinant line bundle $\bigwedge^n (V \otimes \mathcal{L}^k)$ is in some sense optimal. To see this, note
\begin{align*}
    F_{\mathcal{L}} &= \bar{\partial}{\partial} \log H_{\mc{L}} = \frac{1}{m}\bar{\partial}{\partial} \log \det H_{V(k)} \\
    &= \frac{1}{m}\bar{\partial}{\partial} \Tr \log H_{V(k)} = \frac{1}{m}\Tr F_{V(k)}~.
\end{align*}
Let $\lambda^{(k)}$ denote the eigenvalues of $\Lambda F_{V(k)}$, and $\lambda$ be the corresponding eigenvalue of $\Lambda F_V$. Then we see that
\begin{align*}   
\lambda_i &= \lambda_i^{(k)} - k \Lambda F_{\mc{L}} = \lambda_i^{(k)} - \frac{k}{m} \Tr \Lambda F_{V(k)} \\
&= \lambda_i^{(k)} - \frac{k}{m}\sum_j \lambda_j^{(k)}~.
\end{align*}
Choosing $m = k \rank V$, this choice of untwisting has an homogenising effect on the spectrum of $\Lambda F_{V(k)}$ in that it subtracts the arithmetic mean of the eigenvalues at each point on $X$, . This is just as well, because we see that
\begin{equation*}
    \bigwedge^n (V \otimes \mc{L}^k) \simeq (\mc{L}^k)^{\otimes \rank V} \otimes \bigwedge^n V \simeq \mc{L}^{k \rank V}~,
\end{equation*}
as $\det V$ is trivial for an $\textsf{SU}(n)$ bundle. Choosing $m = k \rank V$ is the only choice that yields an honest bundle metric on $\mc{L}$.

\section{Equivariance}\label{app:equivariance}
Recall the cocycle definition of a rank--$r$ complex vector bundle $\mc{V} \rightarrow X$. Let $X = \bigcup_i U_i$ be a finite covering of coordinate charts, together with matrix--valued transition functions on nonempty overlaps
\begin{equation}
    T_{U \rightarrow V}: U \cap V \rightarrow \textsf{GL}(r;\mathbb{C})~.
\end{equation}
These satisfy the cocycle conditions, $T_{U \rightarrow W} = T_{U \rightarrow V} \circ T_{V \rightarrow W}$ on $U \cap V \cap W$. Additionally, $T_{U \rightarrow U} = \mathbf{1}_V$ and $T_{U \rightarrow V}^{-1} = T_{V \rightarrow U}$. A complex vector bundle is defined by a quotient relation on local trivialisations,
\begin{equation}
    \mc{V} = \left( \bigcup_i U_i \times \mathbb{C}^r \right) / \sim~,
\end{equation}
where $(p,s_U) \in U \times \mathbb{C}^r$ is to be identified with $(p,s_V) \in V \times \mathbb{C}^r$ if the sections $s \in \Gamma(X;\mc{V})$ are related by $s_V = T_{U \rightarrow V} s_U$. From this base definition of bundle, various associated bundle constructions may be made. The ones pertinent to our case are:
\begin{itemize}
    \item The dual bundle $\mc{V}^* \rightarrow X$; the rank--$r$ bundle defined by transition data $T_{U \rightarrow V}^{-1}$. Let $\varphi \in \Gamma(X; \mc{V}^*)$ be a section of the dual bundle and $Q = [T_{U \rightarrow V}]$ be the local matrix representation of the transition function. Then by the dual pairing
    \[ \varphi(s) = \varphi^T s = \varphi_ks^k \in C^{\infty}(X;\mathbb{R})~,\]
    which must be independent of the choice of trivialisation, one has the transformation laws
    \begin{equation*}
        s \mapsto Qs\,, \quad \varphi \mapsto (Q^{-1})^T \varphi~.
    \end{equation*}
    In terms of local frames, a local frame $\{e_a\}_{a=1}^r$ for $\mc{V}$ defines a dual frame $\{e^a\}$ for $\mc{V}^*$ via the dual pairing $e^a(e_b) = \delta^a_b$. In particular, the Hermitian metric on $\mc{V}, H \in \Gamma(X; \mc{V}^* \otimes \overline{\mc{V}}^*)$ transforms as
    \[ H \mapsto (Q^{-1})^T H \overline{Q^{-1}} \]
    \item The endomorphism bundle $\mc{V} \otimes \mc{V}^* := \text{End}(\mc{V})$, where $h = h_a^{\phantom{a} b} e^a \otimes e_b$. Locally, this defines a map $h\vert_p = \mc{V}\vert_p \rightarrow \mc{V}\vert_p$ by acting on the right as $s^b \mapsto s^ah_a^{\phantom{a} b}$. By similar considerations to above, the relevant transformation law is
    \begin{equation*}
        s \mapsto Qs\,, \quad h \mapsto (Q^{-1})^T h Q^T~.
    \end{equation*}
\end{itemize}
Since the curvature form $F \in \Omega^{1,1}(\text{End}(\mc{V})$ is valued in $\text{End}(\mc{V})$, it acts on sections as $F: s^a \mapsto s^aF^{\phantom{a} b}_{a \phantom{b} \mu \overline{\nu}}$. It is straightforward to verify the formula for the Chern connection, $F = \bar{\partial}(\partial H H^{-1})$ yields a well--defined section of $\text{End}(\mc{V})$, in the sense that the coordinate representation of the curvature form should transform between trivialisations as (eliding the bundle indices),
\begin{equation}
    (\tilde{F}_{\mu \overline{\nu}}) = (Q^{-1})^T (F_{\mu \overline{\nu}}) Q^T~.
\end{equation}

For the monad bundles examined in Section \ref{sec:examples}, the relevant transition matrices represent the gauge freedom in the definition of the Kodaira embedding $X \hookrightarrow \textsf{Gr}(r; \dim H^0(X;\mc{V}(k)))$, $Q \in \textsf{GL}(r; \mathbb{C})$ \eqref{eq:k_embedding}. By expanding in a suitable basis of sections for the relevant $\mc{V}(k)$, we verify that the respective transformation laws for each section are satisfied up to $O(10^{-5})$ using \texttt{float32} arithmetic. 

\section{Harmonic representative optimisation}\label{app:harmonic_opt_plots}
Here we exhibit the evolution of quantities of interest during optimisation of the harmonic representative as described in Section \ref{sec:physical_yukawa}. Results are reported on three independent runs over an independent validation dataset. Solid lines represent the mean of the runs, and the shaded area shows the $\pm 2 \sigma$ region.

\begin{figure}[htb]
    \centering
    \includegraphics[width=1.0\linewidth]{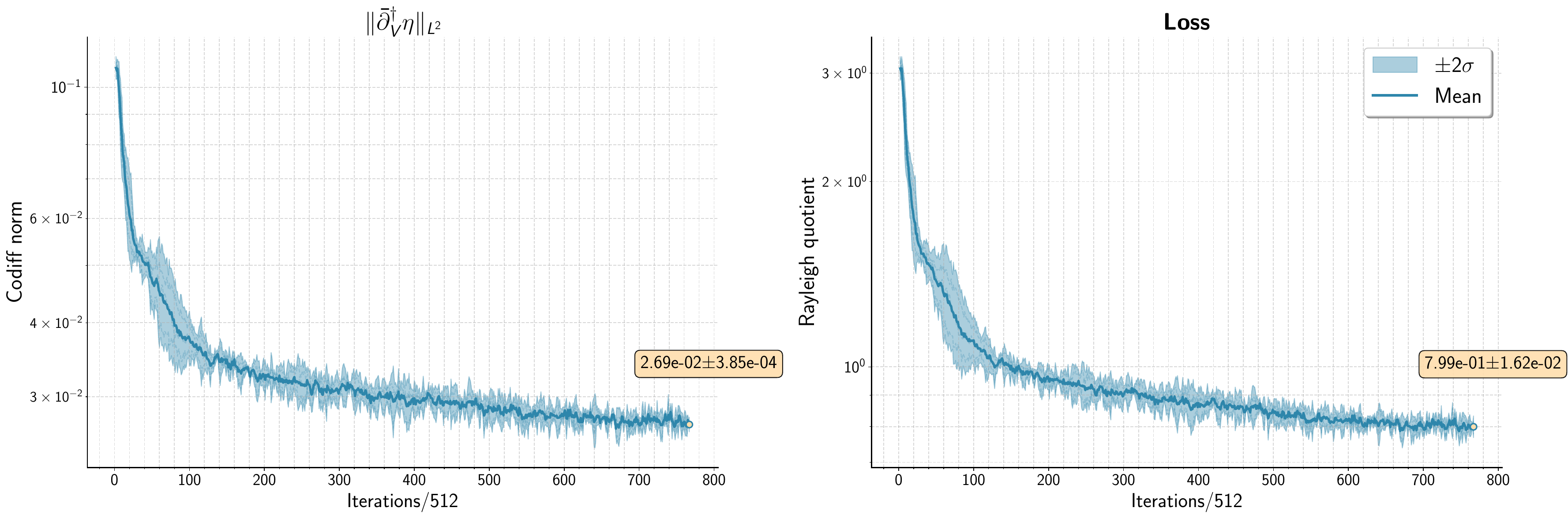}
    \caption{Evolution of the codifferential norm (left) and the variational objective \eqref{eq:rayleigh_quotient} (right) during harmonic representative optimisation for the $\textsf{SU}(3)$ bundle over the Fermat quintic.}
    \label{fig:quintic_AG_harmonic_opt}
\end{figure}
\begin{figure}[htb]
    \centering
    \includegraphics[width=1.0\linewidth]{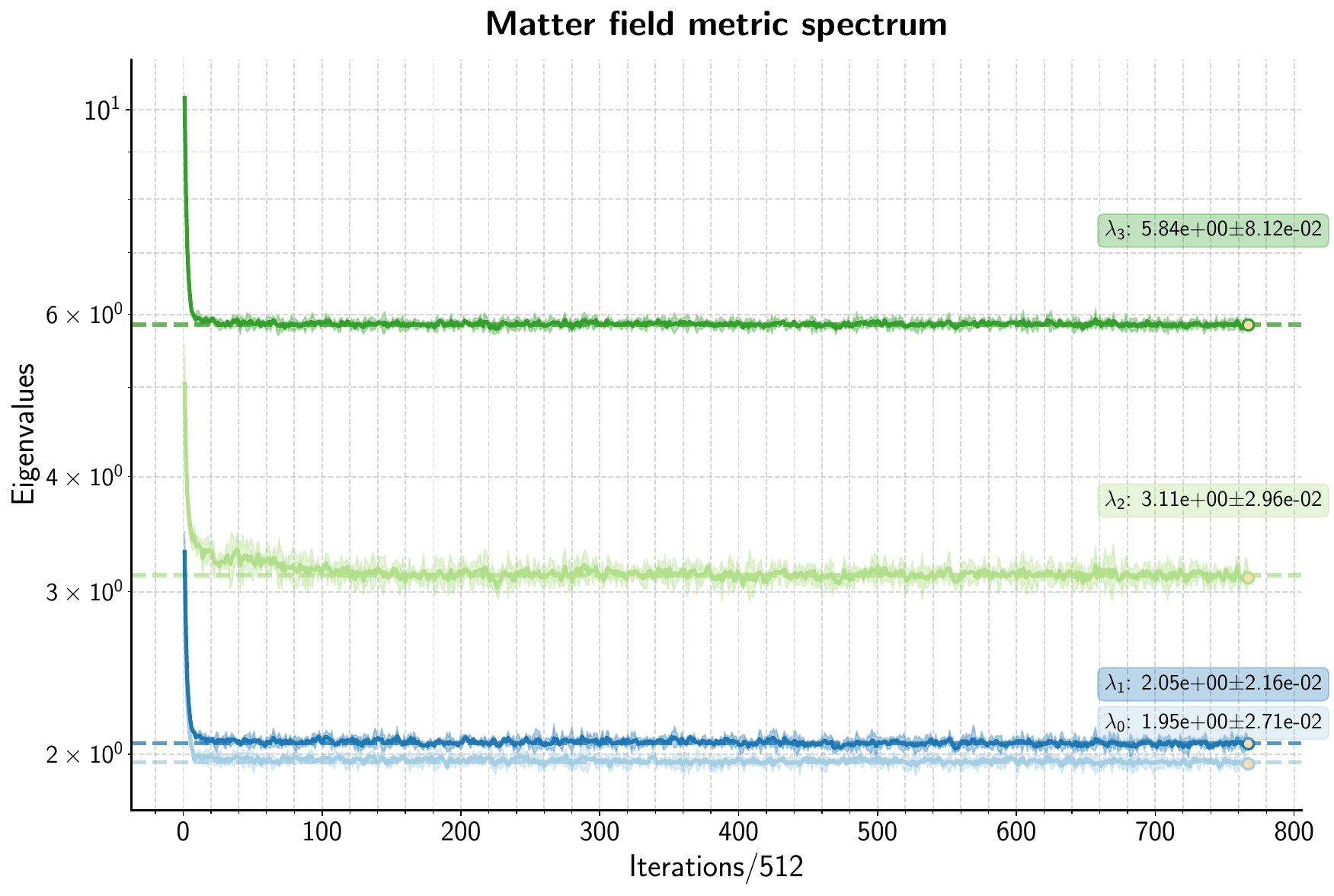}
    \caption{Evolution of the spectrum of the matter K\"ahler field metric $\mathscr{G}$ \eqref{eq:matter_field_metric} during harmonic representative optimisation for the $\textsf{SU}(3)$ bundle over the Fermat quintic, where we vectorise over 4 families simultaneously. Note the rapid convergence of the eigenvalues to their steady--state values after roughly 1000 iterations.}
    \label{fig:G_eigs_4}
\end{figure}
\begin{figure}[htb]
    \centering
    \includegraphics[width=1.0\linewidth]{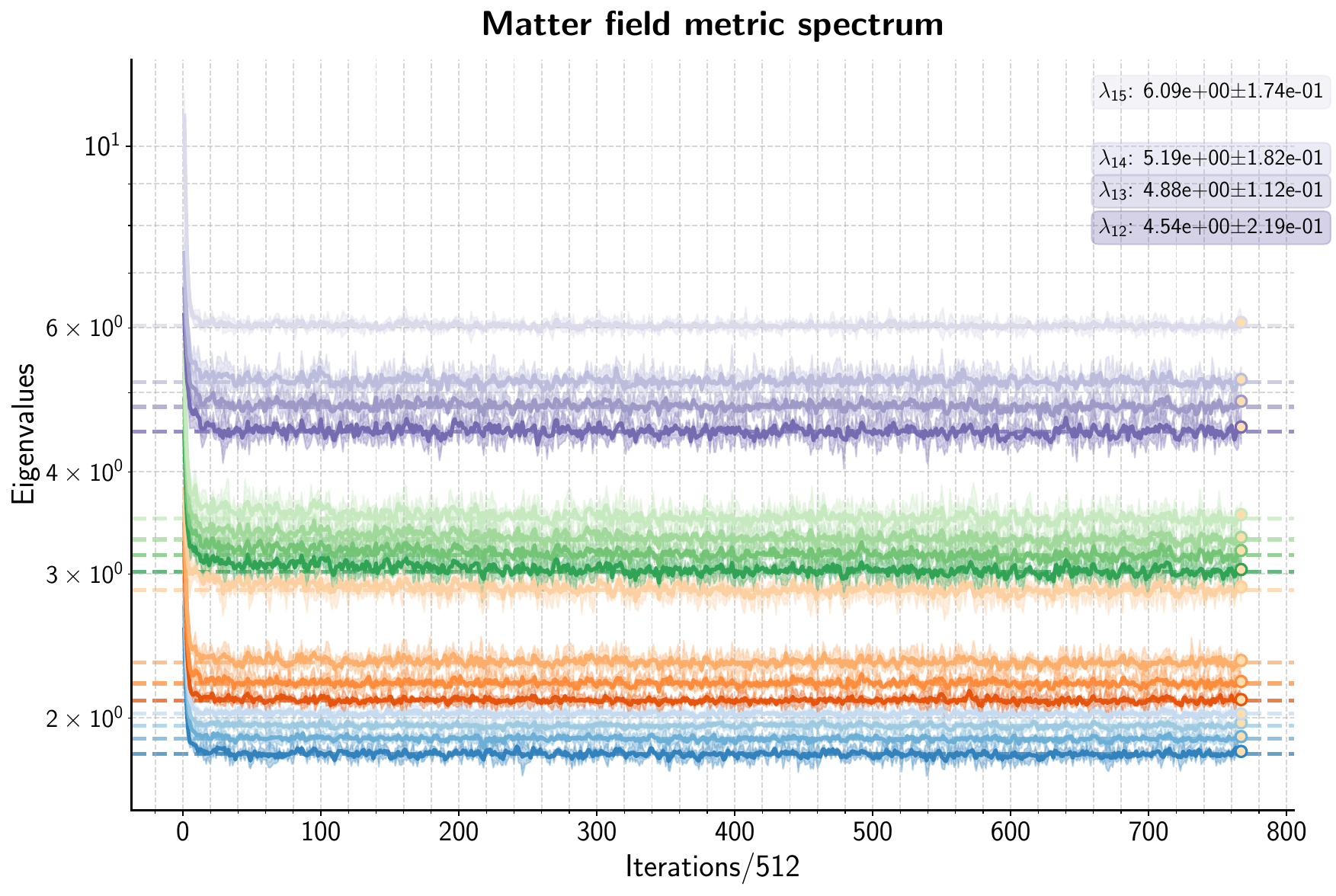}
    \caption{Evolution of the spectrum of the matter K\"ahler field metric $\mathscr{G}$ \eqref{eq:matter_field_metric} during harmonic representative optimisation for the $\textsf{SU}(3)$ bundle over the Fermat quintic, where we vectorise over 16 families simultaneously. Note the rapid convergence of the eigenvalues to their steady--state values after roughly 1000 iterations.}
    \label{fig:G_eigs_16}
\end{figure}

\clearpage

\subsection{Global versus local}\label{app:global_versus_local}
Recall our globally defined ansatz for the harmonic representatives of $H^1(X;V)$ \eqref{eq:harmonic_ansatz}, which we reproduce below
\begin{equation}
    \eta_i = \xi_i + \bar{\partial}_V(\psi_{\theta}^m s_m)~,\, i = 1,\ldots,\dim H^1(X;V)~.
\end{equation}
Here $\psi$ is a global function over $X$, $\psi^m \in C^{\infty}(X;\mathbb{C})$, and the $\{s_m\}$ constitute a basis of sections for $V$. This is manifestly equivariant \textit{w.r.t.} coordinate transformations for both the base $X$ and bundle $V$. We compare this against an ansatz where we have an unconstrained neural network directly model the section $\mathfrak{s} \in \Gamma(V)$,
\begin{equation}
    \eta_i = \xi_i + \bar{\partial}_V(\mathfrak{s}_{\theta})~,\, i = 1,\ldots,\dim H^1(X;V)~.
\end{equation}
We will refer to this as the `local ansatz'. Here, we regard the bundle in terms of distinct local trivialisations --- a densely connected neural network processes the coordinates on $X$ as input and routes the final computation to an `output head' depending on the particular local trivialisation, which is chosen fro numerical stability. Each output head returns $\mathfrak{s}_{\theta}$ expressed in the corresponding local frame --- this is simply a $\rank V$--dimensional vector. To encourage $\mathfrak{s}_{\theta}$ to have the correct transformation properties, we supplement the harmonic objective with a transition loss, to form the total objective;
\begin{equation}
    \mathscr{L}[\mathfrak{s}_{\theta}] = \frac{(\bar{\partial}_V^{\dagger} \eta,\bar{\partial}_V^{\dagger} \eta)}{(\eta, \eta)} + \lambda \cdot \sum_{i \neq j} \Vert \mathfrak{s}_{\theta}^{(j)} - T_{i \rightarrow j} \cdot \mathfrak{s}^{(i)}_{\theta} \Vert_{L^2}~.
\end{equation}
Here $i,j$ index different local frames on $V$ and the transition functions $T_{i \rightarrow j}$ describe how the sections $\mathfrak{s}$ transform between local frames. The Lagrange multiplier $\lambda \in \mathbb{R}^+$ is a constant hyperparameter which should be tuned by hand such that the transition loss scaled by $\lambda$ is commensurate with the harmonic component of the total loss functional. The structure of the loss function is clearly undesirable as it necessitates an essentially arbitrary choice of $\lambda$. More seriously, introducing additional terms to the objective (beyond the original variational functional whose Euler--Lagrange equations constitute the PDE system to be solved) generates spurious local minima, obstructing convergence towards the unique solution $\Delta_V \eta_i =0$. 

\begin{figure}[htb]
    \centering
    \includegraphics[width=1.0\linewidth]{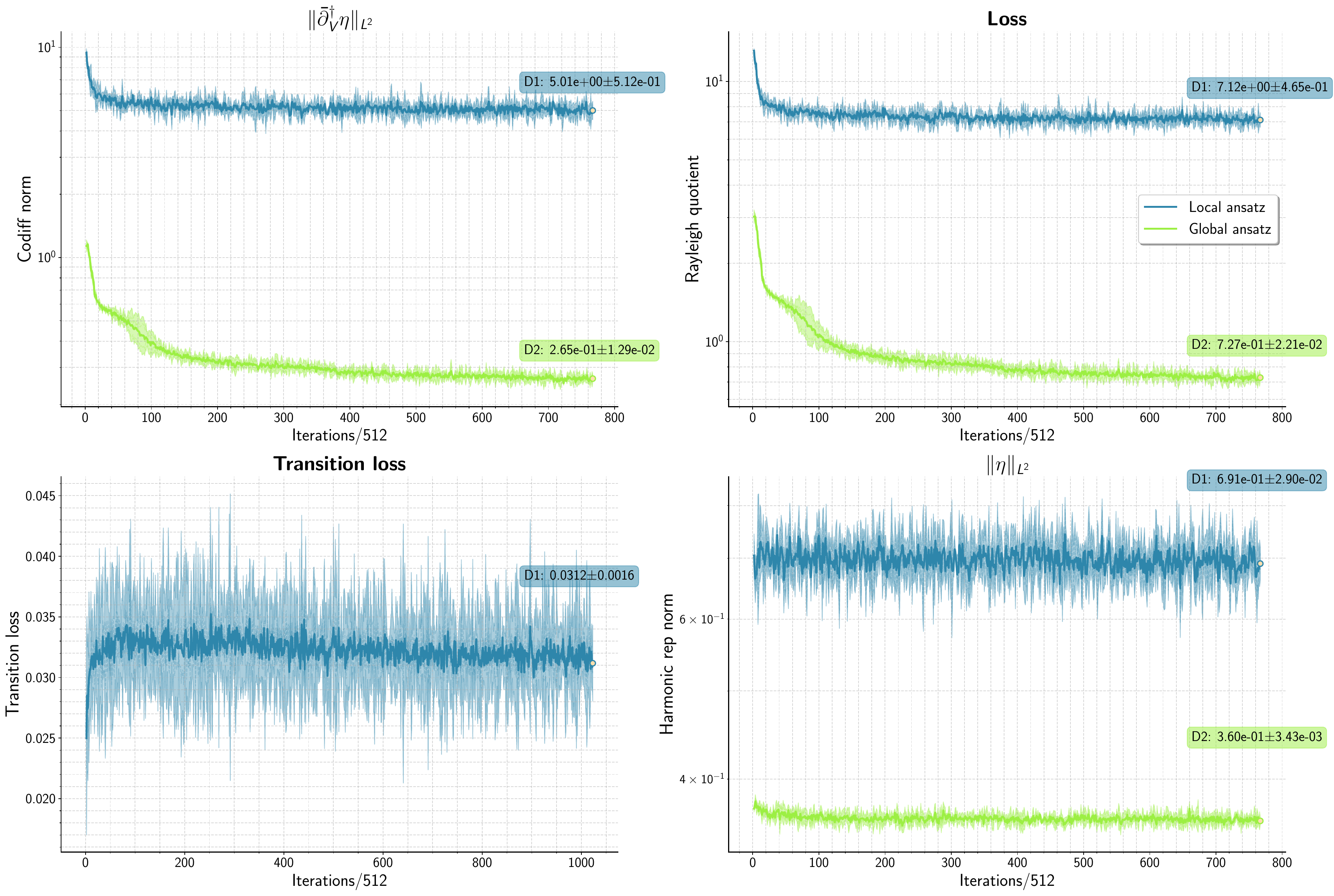}
    \caption{Evolution of various quantities during harmonic representative optimisation for the $\textsf{SU}(3)$ bundle over the Fermat quintic, where we vectorise over four families simultaneously. We elide the transition loss for the global ansatz as this is satisfied by construction. Note the significantly reduced variance for each quantity exhibited by the global ansatz.}
    \label{fig:global_v_local}
\end{figure}

We compare both the global and local models by optimising for the harmonic representatives of four distinct classes in $H^1(X;V)$ and exhibit the results in Figure \ref{fig:global_v_local}. We logarithmically scan through $\lambda \in [0.1,100]$  to identify the value of $\lambda$ that leads to the lowest value of the codifferential norm at the end of optimisation. Immediately we see that the global model is able to achieve a significantly lower final value of the norm of the codifferential $\bar{\partial}_V^{\dagger} \eta$ and exhibits significantly reduced variance from the local model in general. This suggests that the local model is only capable of producing a collection of local fields $\{\mathfrak{s}_{\theta}^{(i)}\}_i$ which cannot be coherently patched together into a globally defined field.

We see that the topology of any nontrivial manifold represents a `soft' topological obstruction to learning. If our parameterised function class and optimiser had infinite modelling capacity, in principle there is nothing preventing us from converging to the unique harmonic representative without a manifestly equivariant architecture. This is not true in general. In this example, for the optimisation procedure to reliably converge to the desired result, we must embed the desired mathematical structure into our ansatz by construction to minimise the space of possible hypotheses that must be traversed by the optimiser. This is the simple thesis of geometric machine learning. 

\clearpage
\section{Visualisations of couplings}\label{app:coupling_vis}
Here we present further visualisations of the holomorphic and physically normalised couplings computed using the procedure outlined in Section \ref{sec:physical_yukawa}. We repeat the exercise for a subset of 9, 16 and 25 families represented by distinct cohomology classes in $H^1(X;V)$, which is of dimension 50. We may efficiently obtain the harmonic representatives for multiple cohomology classes within a single optimisation procedure by vectorisation over the family index.
\begin{figure}[htb]
    \centering
    \includegraphics[width=1.0\linewidth]{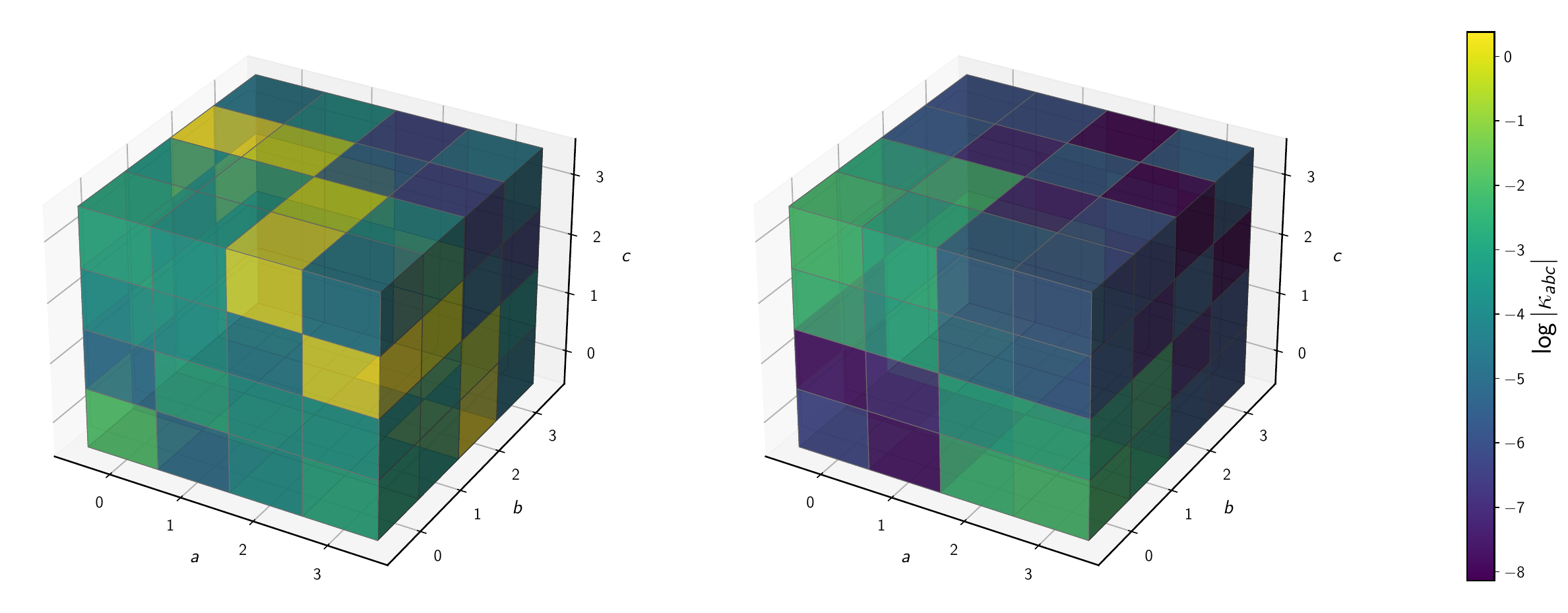}
    \caption{Voxel plots of the holomorphic (left) and normalised (right) couplings $\kappa_{abc}$ for four distinct classes in the cohomology $H^1(X;V)$.}
    \label{fig:voxel_compare_couplings}
\end{figure}
\begin{figure}[htb]
    \centering
    \includegraphics[width=1.0\linewidth]{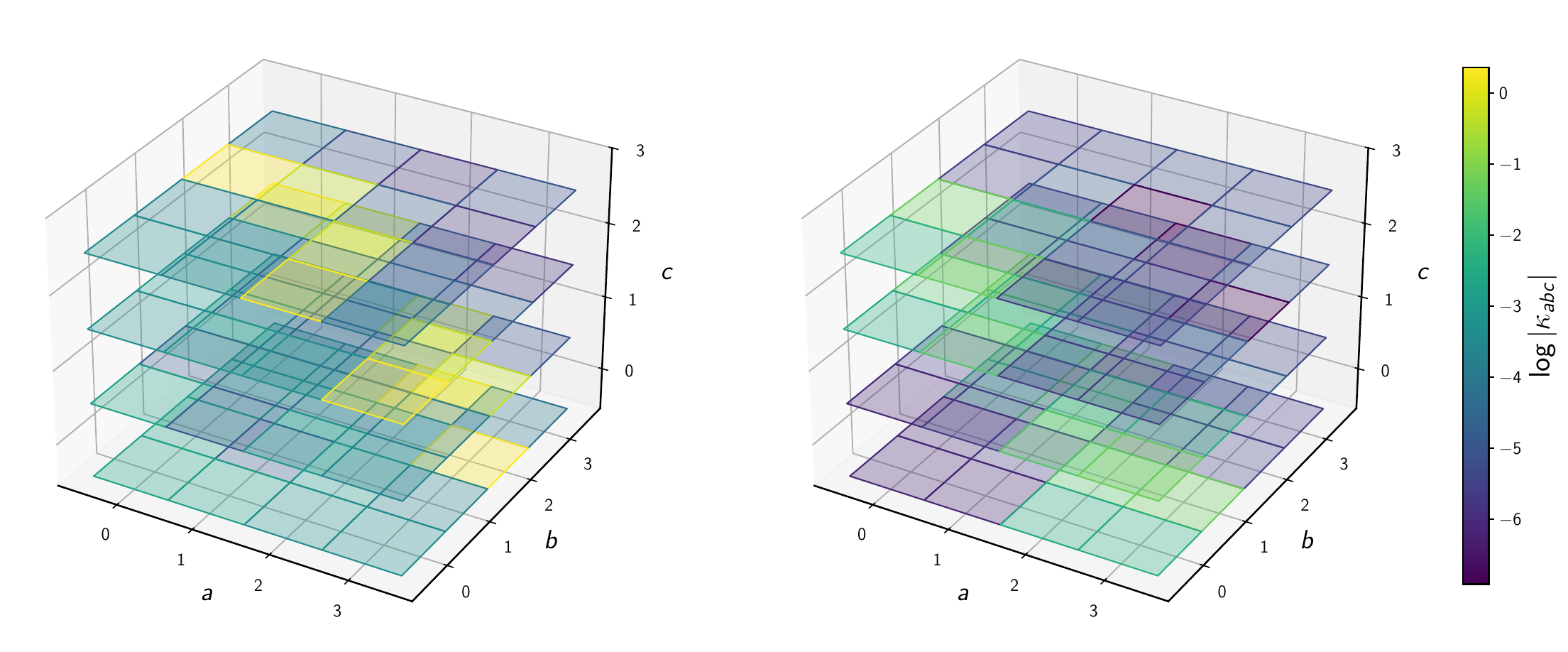}
    \caption{Voxel plots, sliced along the $Z$ axis for the holomorphic (left) and normalised (right) couplings $\kappa_{abc}$ for four distinct classes in the cohomology $H^1(X;V)$.}
    \label{fig:voxel_slices_compare_couplings}
\end{figure}
\begin{figure}[htb]
    \centering
    \includegraphics[width=1.0\linewidth]{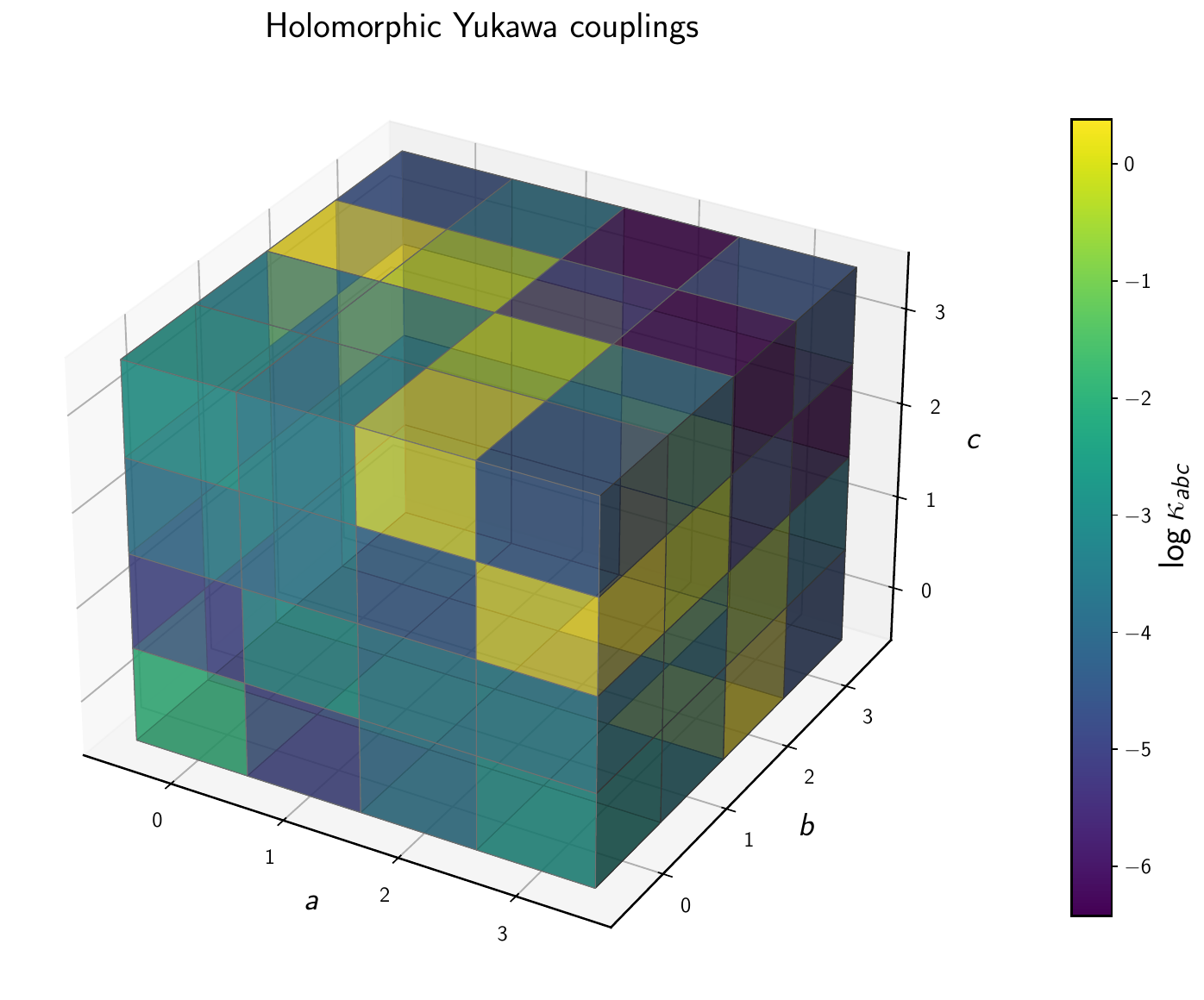}
    \caption{Voxel plot for the holomorphic couplings $\kappa_{abc}$ for four distinct classes in the cohomology $H^1(X;V)$.}
    \label{fig:voxel_slices_compare_couplings4}
\end{figure}
\begin{figure}[htb]
    \centering
    \includegraphics[width=1.0\linewidth]{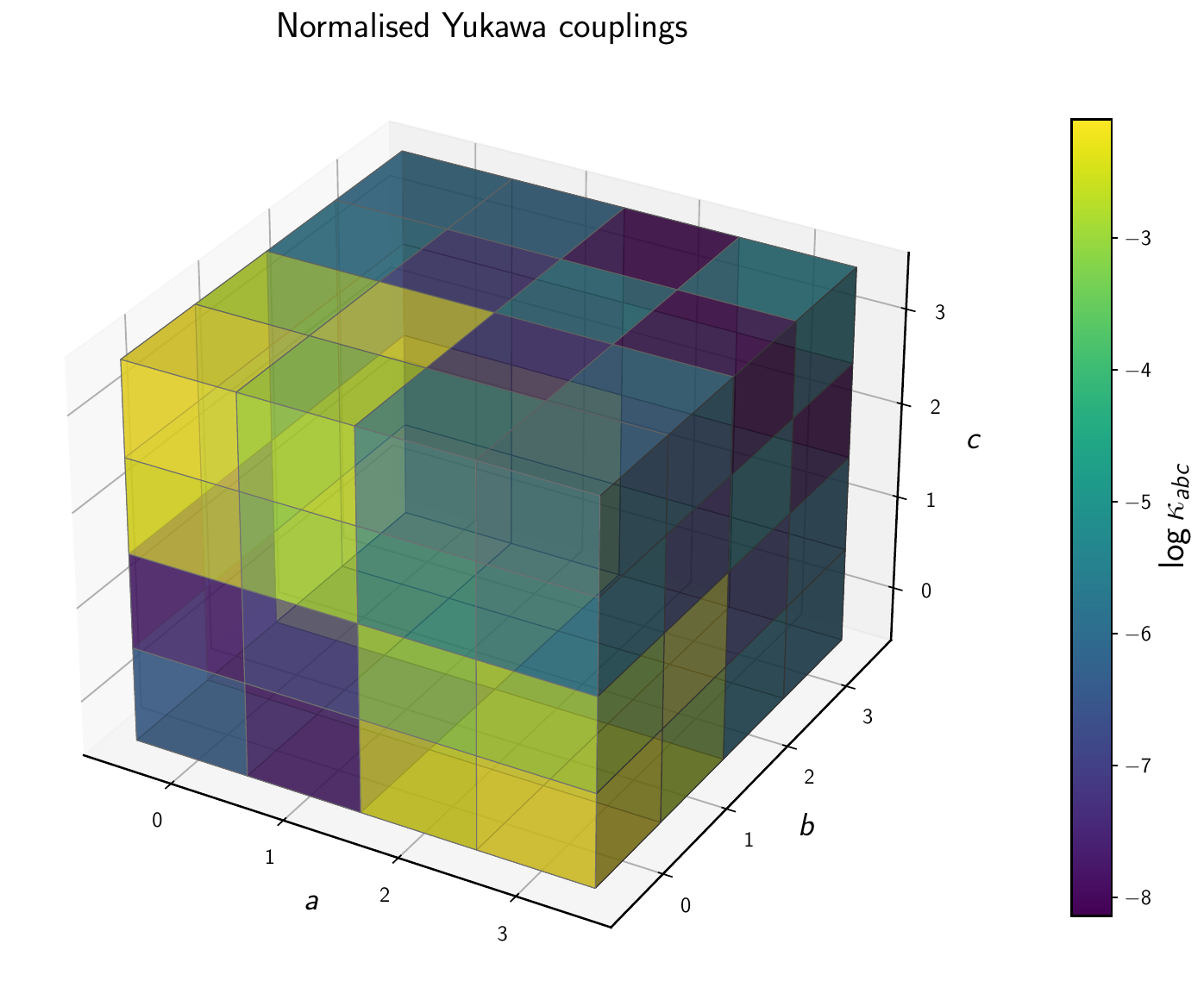}
    \caption{Voxel plot for the normalised couplings $\widetilde{\kappa}_{abc}$ for four distinct classes in the cohomology $H^1(X;V)$.}
    \label{fig:voxel_slices_compare_couplings4_norm}
\end{figure}

\begin{figure}[htb]
    \centering
    \includegraphics[width=1.0\linewidth]{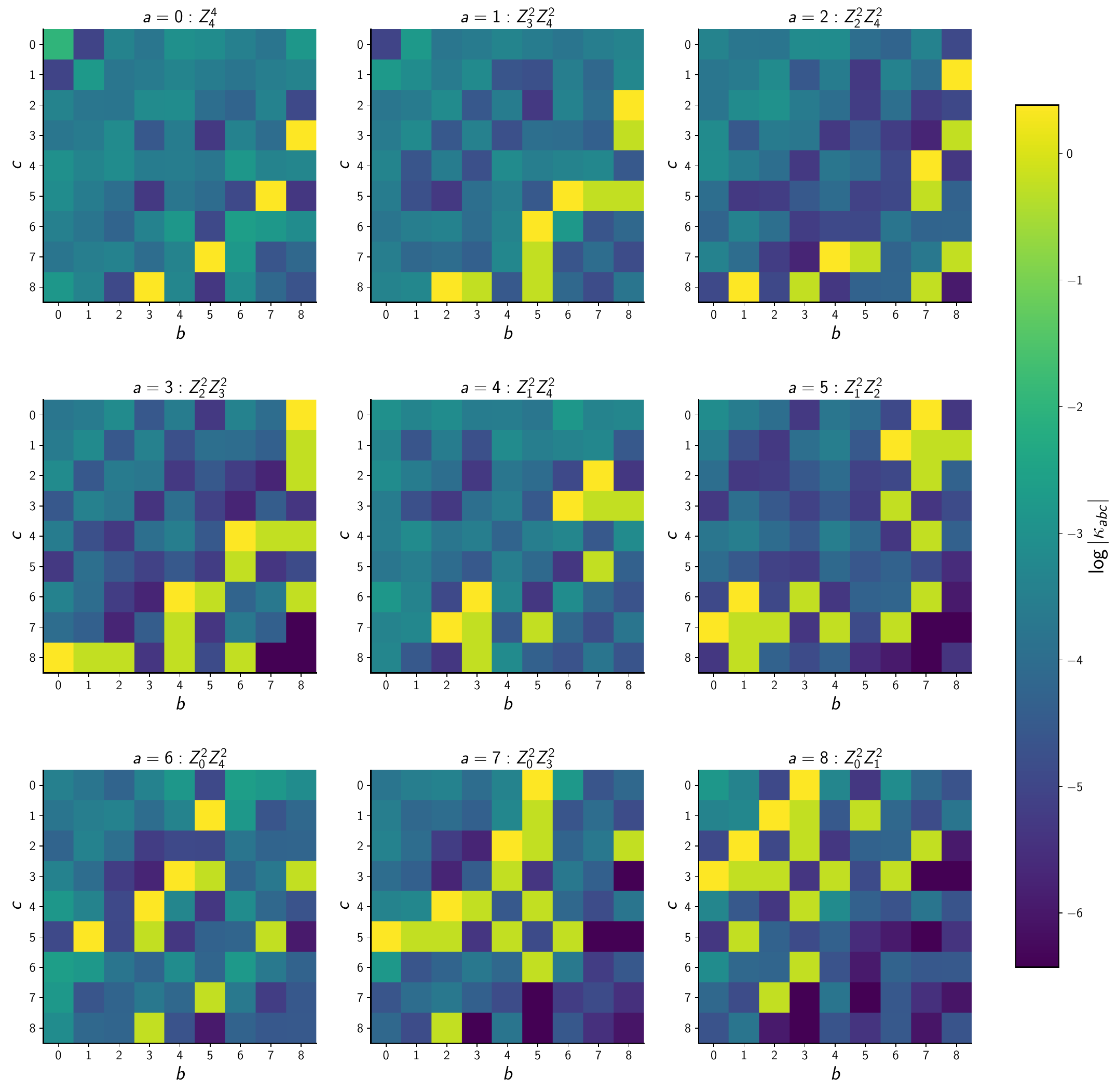}
    \caption{Heatmap of slices of the holomorphic Yukawa coupling array $\kappa_{abc}$ for nine distinct classes in the cohomology $H^1(X;V)$. These classes may be represented algebraically using the indicated polynomials. This computation is semi--analytic.}
    \label{fig:holo_couplings_big}
\end{figure}

\begin{figure}[htb]
    \centering
    \includegraphics[width=1.0\linewidth]{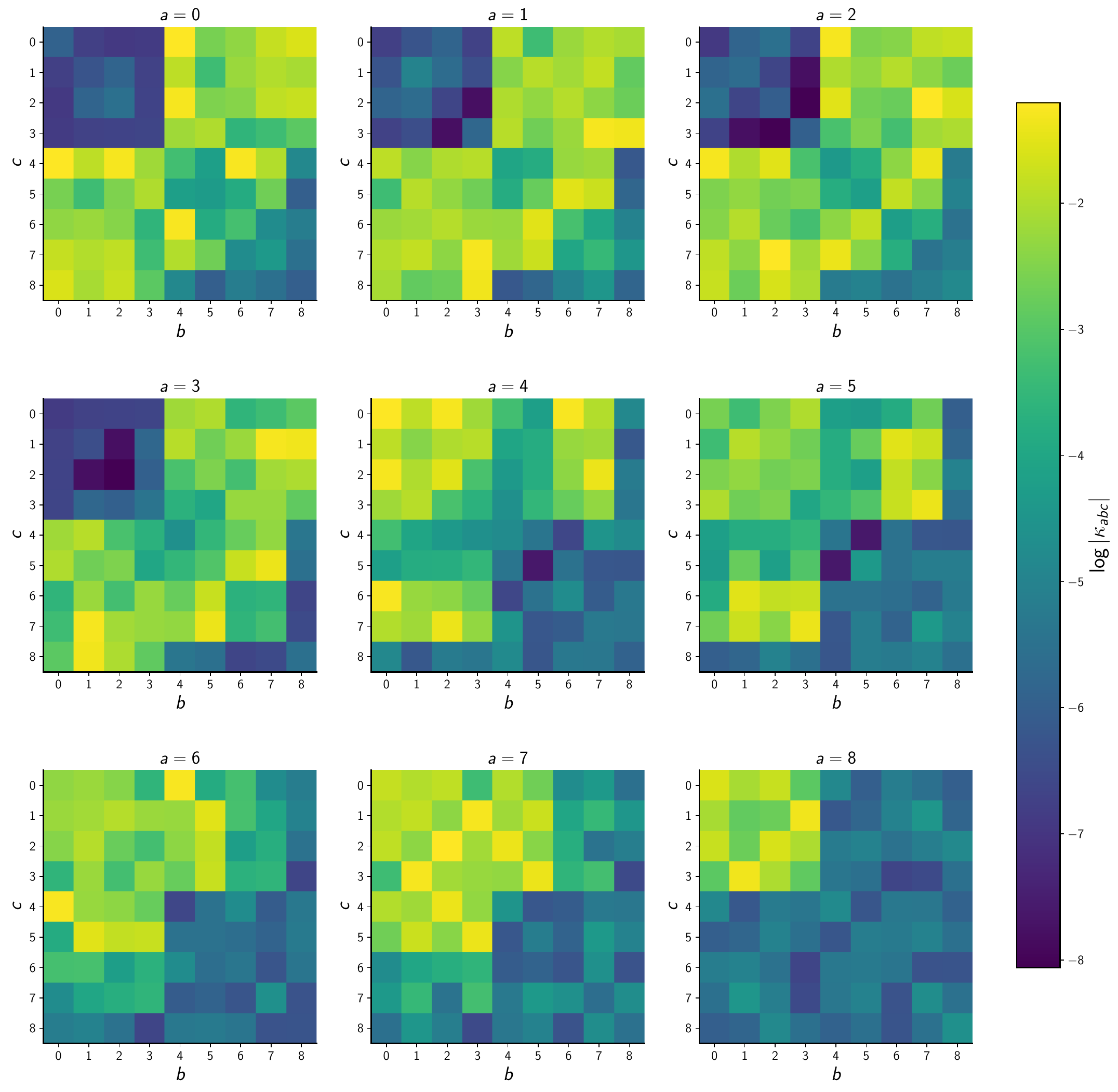}
    \caption{Heatmap of slices of the normalised Yukawa coupling array $\widetilde{\kappa}_{abc}$, evaluated in the eigenbasis of the matter field metric $\mathscr{G}_{a\overline{b}}$ \eqref{eq:matter_field_metric}. This metric is evaluated using nine distinct classes in the cohomology $H^1(X;V)$. This computation is differential--geometric and depends on our learned approximations.}
    \label{fig:norm_couplings_big}
\end{figure}

\begin{figure}[htb]
    \centering
    \includegraphics[width=1.0\linewidth]{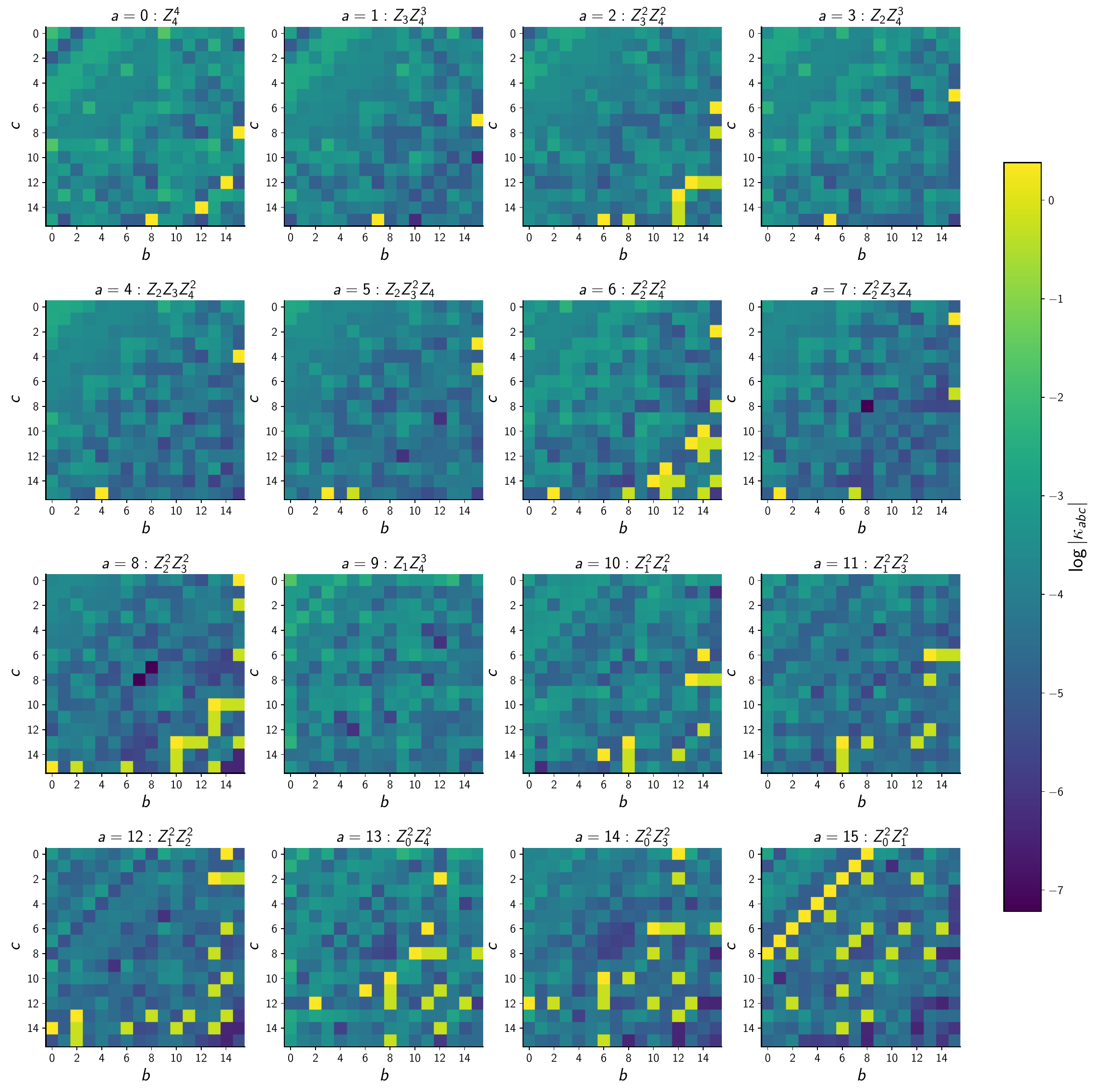}
    \caption{Heatmap of slices of the holomorphic Yukawa coupling array $\kappa_{abc}$ for 16 distinct classes in the cohomology $H^1(X;V)$. These classes may be represented algebraically using the indicated polynomials. This computation is semi--analytic.}
    \label{fig:holo_couplings_big}
\end{figure}

\begin{figure}[htb]
    \centering
    \includegraphics[width=1.0\linewidth]{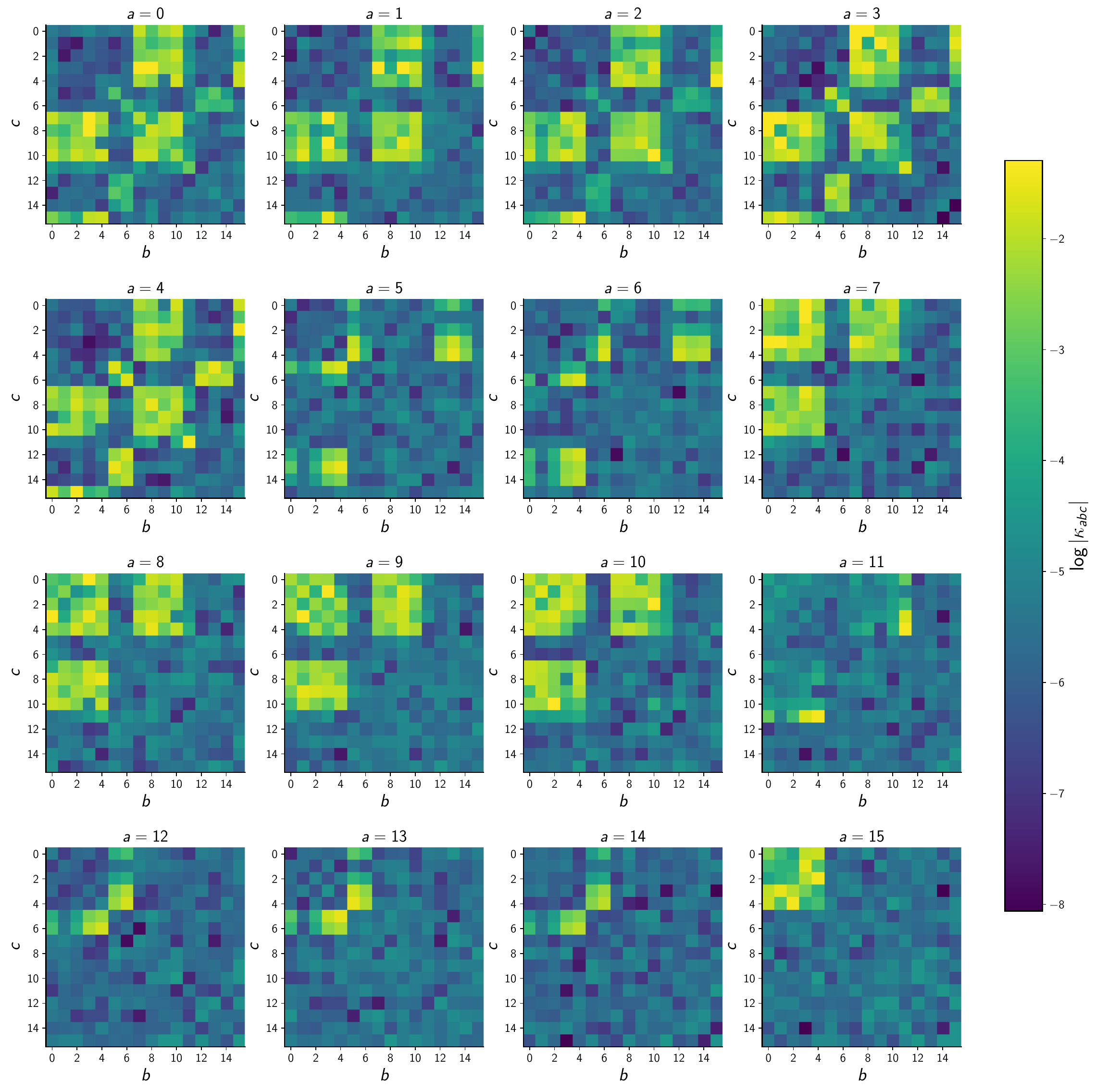}
    \caption{Heatmap of slices of the normalised Yukawa coupling array $\widetilde{\kappa}_{abc}$, evaluated in the eigenbasis of the matter field metric $\mathscr{G}_{a\overline{b}}$ \eqref{eq:matter_field_metric}. This metric is evaluated using 16 distinct classes in the cohomology $H^1(X;V)$. This computation is differential--geometric and depends on our learned approximations.}
    \label{fig:norm_couplings_big}
\end{figure}

\begin{figure}[htb]
    \centering
    \includegraphics[width=1.0\linewidth]{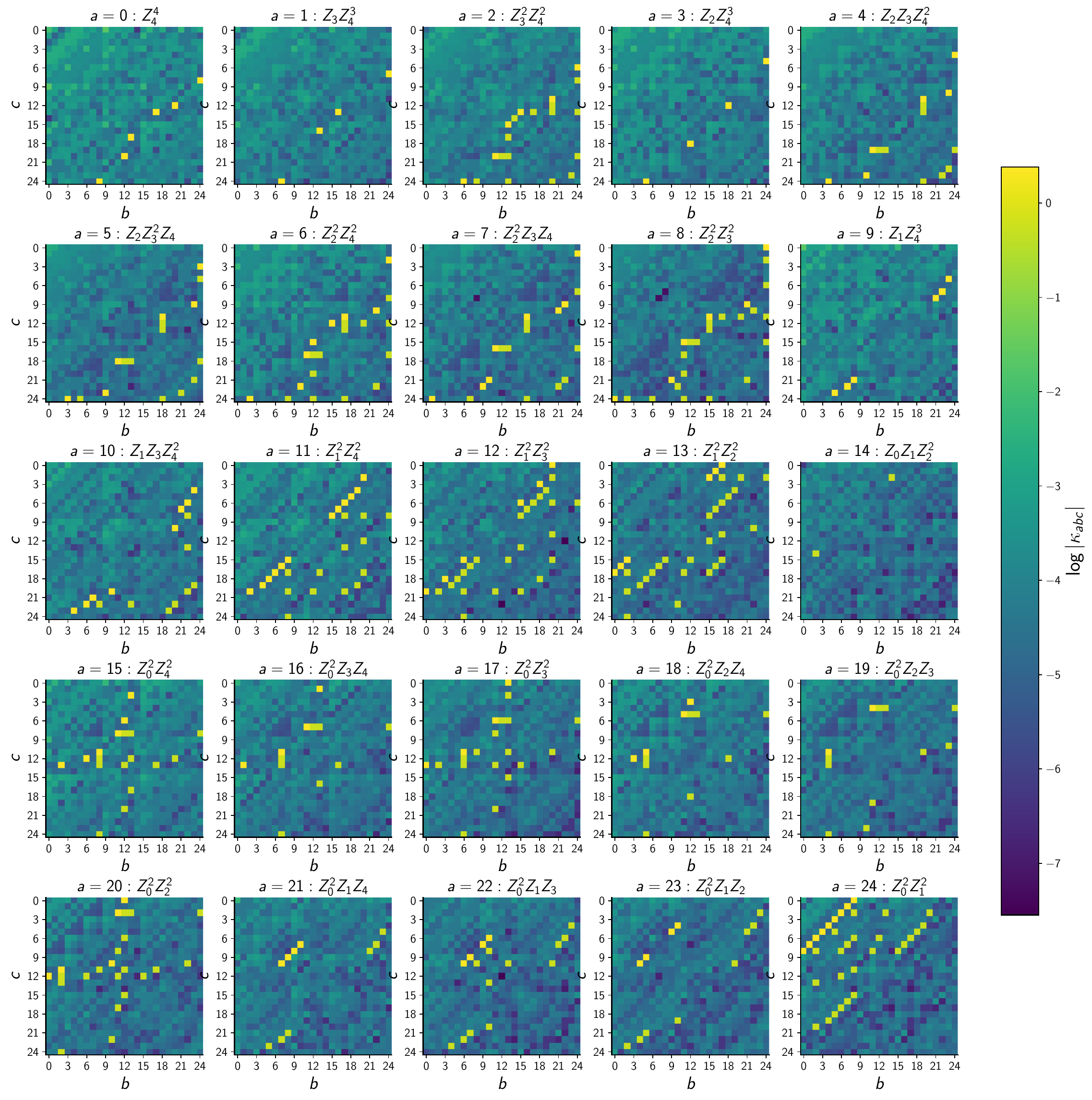}
    \caption{Heatmap of slices of the holomorphic Yukawa coupling array $\kappa_{abc}$ for 25 distinct classes in the cohomology $H^1(X;V)$. These classes may be represented algebraically using the indicated polynomials. This computation is semi--analytic.}
    \label{fig:holo_couplings_big}
\end{figure}

\begin{figure}[htb]
    \centering
    \includegraphics[width=1.0\linewidth]{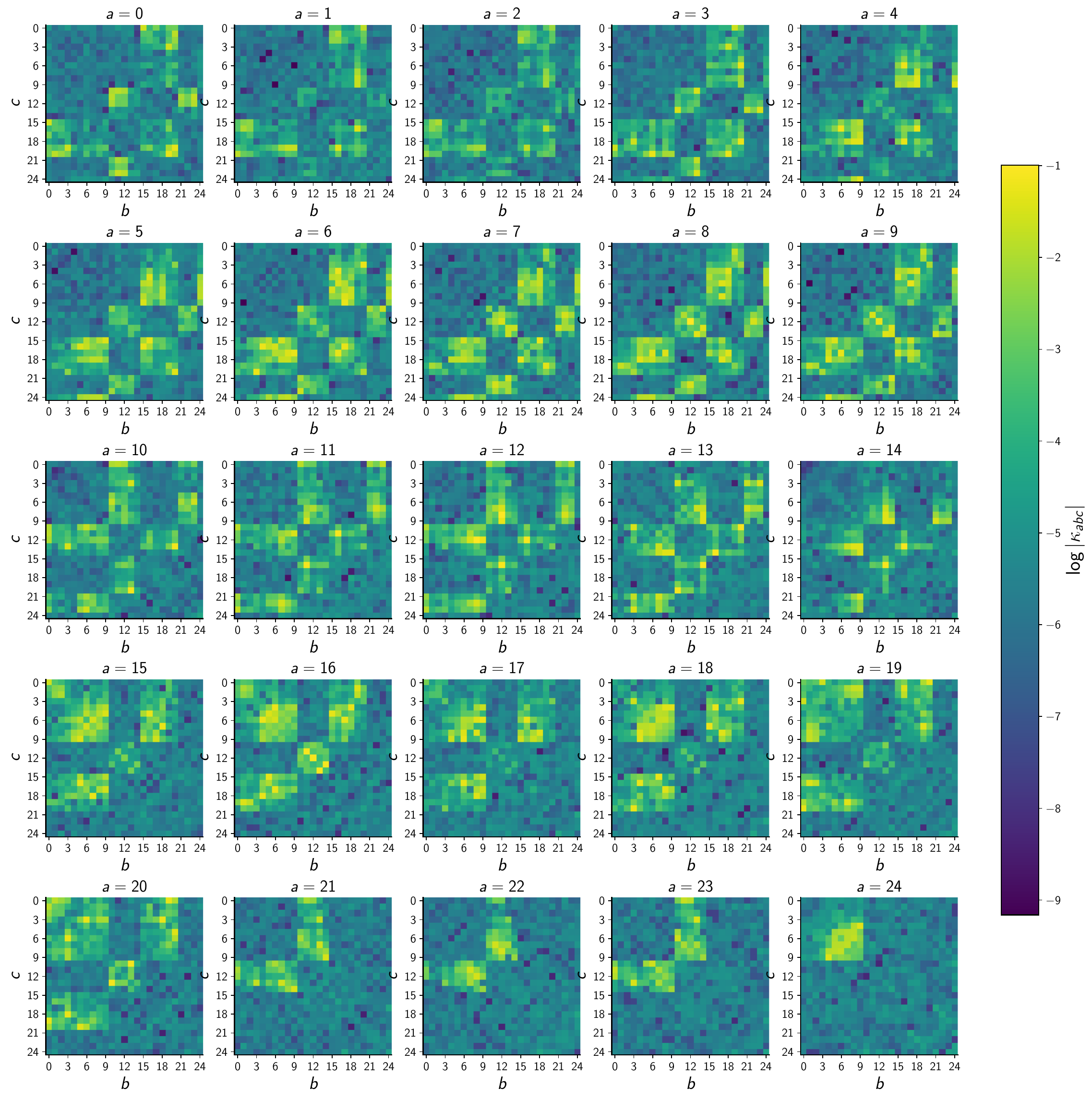}
    \caption{Heatmap of slices of the normalised Yukawa coupling array $\widetilde{\kappa}_{abc}$, evaluated in the eigenbasis of the matter field metric $\mathscr{G}_{a\overline{b}}$ \eqref{eq:matter_field_metric}. This metric is evaluated using 25 distinct classes in the cohomology $H^1(X;V)$. This computation is differential--geometric and depends on our learned approximations.}
    \label{fig:norm_couplings_big}
\end{figure}

\end{appendices}

\end{document}